\documentclass{article}
\usepackage[utf8]{inputenc}
\usepackage{jcappub}
\usepackage{aas_macros}
\usepackage{float}
\usepackage{amsmath,amsfonts,amssymb}
\usepackage{mathrsfs}
\usepackage{graphicx}
\usepackage[english]{babel} 
\usepackage{hyperref}
\usepackage[dvipsnames]{xcolor}
\usepackage{adjustbox}
\usepackage[normalem]{ulem}
\usepackage[capitalize]{cleveref}
\usepackage{csquotes}
\usepackage{multirow}

\newcommand{\element}[2]{\ensuremath{{}^{#1}\mathrm{#2}}}
\newcommand{\newtabline}{\hline \rule{0pt}{1em}\hspace*{0pt}}

\newcommand{\nils}[1]{{\color{black}#1}}
\newcommand{\nilsr}[2]{#2}
\title{The 2024 BBN baryon abundance update}
\date{January 2024}

\author[1]{Nils Sch\"oneberg,}
\affiliation[1]{Institut de Ci\`encies del Cosmos, Universitat de Barcelona, Mart\'{\i} i Franqu\`es 1, Barcelona 08028, Spain}

\abstract{We revisit the state of the light element abundances from big bang nucleosynthesis in early 2024 with particular focus on the derived baryon abundance. We find that the largest differences between the final baryon abundances are typically driven by the assumed Deuterium burning rates, characterized in this work by the underlying code. The rates from theoretical ab-initio calculations favor smaller baryon abundances, while experimentally-determined rates prefer higher abundances. Through robust marginalization over a wide range of nuclear rates, the recently released \texttt{PRyMordial} code allows for a conservative estimate of the baryon abundance at $\Omega_b h^2 = 0.02218 \pm 0.00055$ (using PDG-recommended light element abundances) in $\Lambda$CDM and $\Omega_b h^2 = 0.02196 \pm 0.00063$ when additional ultra-relativistic relics are considered ($\Lambda$CDM + $N_\mathrm{eff}$). These additional relics themselves are constrained to $\Delta N_\mathrm{eff} = -0.14 \pm 0.21$ by light element abundances alone.}

\begin{document}

\emailAdd{nils.science@gmail.com}

\maketitle

\section{Introduction}

The abundance of baryons in the universe is one of the most important but least discussed ingredients of the $\Lambda$CDM model. Indeed, it is critical for almost any discussion of its thermal history, playing a role throughout the big bang nucleosynthesis (BBN), the primordial baryonic acoustic oscillations (BAO), the release of the cosmic microwave background (CMB), and the formation of the earliest stars and galaxies. As such, a precise knowledge of this fundamental parameter of any cosmological model is absolutely crucial for almost all predictions. Yet, compared to more exotic parameters such as the dark matter (or total matter) abundance or the dark energy characteristics, this parameter is comparatively less discussed.

While the parameter is well constrained from the CMB observations \cite{Planck:2018vyg}, in view of the emerging tensions such as the Hubble tension (now reaching beyond $5 \sigma$ significance) \cite{Riess:2021jrx,Verde:2023lmm,Freedman:2023jcz}, it is crucial to find independent internal consistency checks for this model parameter. The light element abundances that were generated during BBN offer such a crucial consistency check \cite{Pisanti:2020efz,Yeh:2020mgl,Pitrou:2020etk}. As such, this short update/review paper is dedicated to presenting an approximate snapshot from early 2024 of the current state of the baryon abundance determinations from BBN (both from the theoretical and the experimental side), as well as a brief pedagogic review of the importance of the baryon abundance in cosmology.

The issue of the baryon abundance is particularly timely now due to the advent of high-precision surveys of the CMB (ACT, SPT, CMB-S4) and of the large-scale structure (Rubin, Euclid, DESI, etc.). Both of these cosmological probes are fundamentally connected to the acoustic oscillations of the baryons, which imprint both as the main oscillations of the CMB and as oscillations and shape of the large scale structure power spectrum.\footnote{This shape and the corresponding slope at the turnover result from the damped growth of the matter fluctuations on scales where the baryonic acoustic oscillations have prevented a clustering of the baryons similar to that of cold dark matter.} The speed and the corresponding wavelength of the BAO (the sound horizon) is directly related to the baryon abundance. As such, the measurements of the baryon abundance play a crucial role in calibrating the sound horizon standard ruler, allowing the BAO to be used as a probe of the Hubble constant (see for example \cite{Addison:2013haa,Aubourg:2014yra,Addison:2017fdm,Blomqvist:2019rah,Cuceu:2019for,Schoneberg:2019wmt,Schoneberg:2022ggi}). 

Similarly, the suppression of the power spectrum related to the ratio of abundances of baryonic and cold dark matter can be used as a further probe of cosmology once calibrated. This is what gives the approaches fitting the full power spectrum (full-modeling \cite{Ivanov:2019pdj,DAmico:2019fhj,Philcox:2020vvt,Chudaykin:2020aoj,Wadekar:2020hax,Kobayashi:2021oud,Chen:2021wdi,Smith:2022iax,Simon:2022lde,Holm:2023laa}) as well as the ShapeFit approach (see \cite{Brieden:2021cfg,Brieden:2021edu,Brieden:2022lsd}) their constraining power. While in future surveys there is a distinct possibility of determining the baryon abundance from these large-scale structure surveys themselves, the abundance as derived from the BBN serves now and will continue to serve as a crucial cross-check. Furthermore, with recent theoretical and experimental advances, predictions of the light element abundances from BBN are now more accurate than ever before.

We give a short pedagogic summary of a few theoretical aspects in \cref{sec:theory}, describe the selection of codes and data used in this work in \cref{sec:method_data}, summarize our results in \cref{sec:results}, and conclude in \cref{sec:conclusions}. Any reader versed in the theoretical aspects of BBN can safely skip to \cref{sec:method_data}.

\section{A short theory summary} \label{sec:theory}

The aim of this section is not to review all of BBN -- we leave this task to excellent reviews such as \cite{Grohs:2023voo,Cyburt:2015mya} or the works describing the respective codes such as \cite{Pitrou:2018cgg,Consiglio:2017pot,Pisanti:2020efz}. Instead, this section aims to give a short description of the most important aspects allowing the reader to quickly develop a physical intuition of how the different parts of BBN work and how the abundances are connected.

After the primordial quark-gluon plasma condensed into protons and neutrons (the only relatively stable baryons), the universe is dominated by a bath of photons and neutrinos, electrons and positrons, and protons and neutrons. Importantly, due to baryogenesis, the number of photons is around $10^{9}$ times larger than the number of baryons ($\eta_b = n_b/n_\gamma \approx 6 \cdot 10^{-10}$).

At around $T \simeq 0.7\mathrm{MeV}$ almost simultaneously the electrons and positrons begin their annihilation (heating up the photons and neutrinos) and the weak reactions freeze out. The freeze-out of the weak reactions decouples the neutrinos from the primordial plasma and prevents further proton-neutron conversion reactions (such as $\nu_e + n \leftrightarrow e^- + p^+$ or $n + e^+ \leftrightarrow \bar{\nu}_e + p^+$), causing a freeze-out also of the neutron and proton abundance --- up to the residually allowed (but much slower) $\beta^-$ decay ($n \rightarrow p^+ + e^- + \bar{\nu}_e$). As such, after this threshold is reached, the neutrons start slowly decaying with an initial abundance given approximately by their Boltzmann factor ($n_n/n_p \simeq \exp(-(m_n-m_p)/T) \simeq 0.15$) until they are fused into the heavier elements.

One could expect that any heavier element could easily be fused by just combining enough protons and neutrons, but this is not true. The fusion of heavier elements is suppressed by higher powers of the tiny baryon-to-photon ratio ($\eta_b$), and thus the elements always have to be built up in subsequent fusions from already existing lighter elements. Given that the fusion into Deuterium only happens relatively late (see below), this 'Deuterium bottleneck' is the fundamental reason why so much primordial matter remains in relatively light elements.

Because the Deuterium binding energy is around 2.22MeV it may seem contradictory that the fusion of such Deuterium nuclei begins only long after the weak freeze-out threshold at $0.7\mathrm{MeV}$. The solution to this puzzle is that the fusion of the light nuclei is constantly opposed by photo-disintegration, and since the photons outnumber the baryons one to $10^{9}$, the Deuterium nuclei can still be disintegrated by high-energy photons in the high-frequency tail of the photon phase-space distribution, even when the bulk of photons carry an energy much too small to do that. This fact is enough to delay the fusion of Deuterium nuclei until the mean photon energy is around $T \simeq 0.07\mathrm{MeV}$.\enlargethispage*{3\baselineskip}

As such, Deuterium fusion only begins relatively late. At that point further fusion to heavier elements (like Tritium or Helium) is highly energetically beneficial, and the Deuterium is almost immediately fused away.\footnote{The mass gaps at mass numbers $A=5$ and $A=8$ with the unstable elements $\element{5}{He}$ or $\element{5}{Li}$ and $\element{8}{Be}$ conspire to make the fusion of even heavier stable elements during BBN extremely unlikely, requiring stars to generate enough heat for the unlikely triple-alpha process to create these heavier elements.} This immediate process of converting Deuterium to heavier elements like Tritium and Helium-3 (often dubbed 'Deuterium-burning') is the reason why the remaining abundance of Deuterium in the primordial plasma is only around $10^{-5}$ relative to the Hydrogen. This represents a small enough abundance such that further fusion reactions are too inefficient to happen frequently. 

The efficiency of heavier fusion reactions means that almost all neutrons that are initially used up for Deuterium are further fused into $\element{4}{He}$. Indeed, a simple estimate can be made that all neutrons are used up in the Helium production, leading to a fraction of around $Y_p \simeq 0.24$ (which is very close to actual predictions).

\noindent We can thus summarize as follows:
\begin{itemize}
    \item The formation of the elements is delayed to $T \simeq 0.07\mathrm{MeV}$ due to the small baryon-to-photon ratio $\eta_b$ and begins with the formation of Deuterium.
    \item Deuterium is almost immediately burnt up, with the reaction rates of these fusion reactions being fundamentally important for the final deuterium abundance (see \cref{sec:method_data}). Since the start of the Deuterium burning is tightly connected to the baryon-to-photon ratio, the Deuterium abundance is also very tightly connected to the baryon abundance (more baryons $\to$ higher temperature Deuterium fusion $\to$ stronger Deuterium burning $\to$ lower final Deuterium abundance).
    \item The final Helium abundance is mostly determined by the time that the neutrons have to decay between the freeze-out of weak reactions and the start of nucleosynthesis (as well as the nuclear rates). As such, the abundance is very tightly connected to the overall Hubble expansion rate (connecting physical time to temperature energy scales), which at this early time is given directly by the neutrino number $\Delta N_\mathrm{eff}$ (higher $N_\mathrm{eff}$ $\to$ higher Hubble rate $\to$ younger universe $\to$ more neutrons $\to$ more Helium).
\end{itemize}

\begin{figure}[ht]
    \centering
    \includegraphics[width=0.75\textwidth]{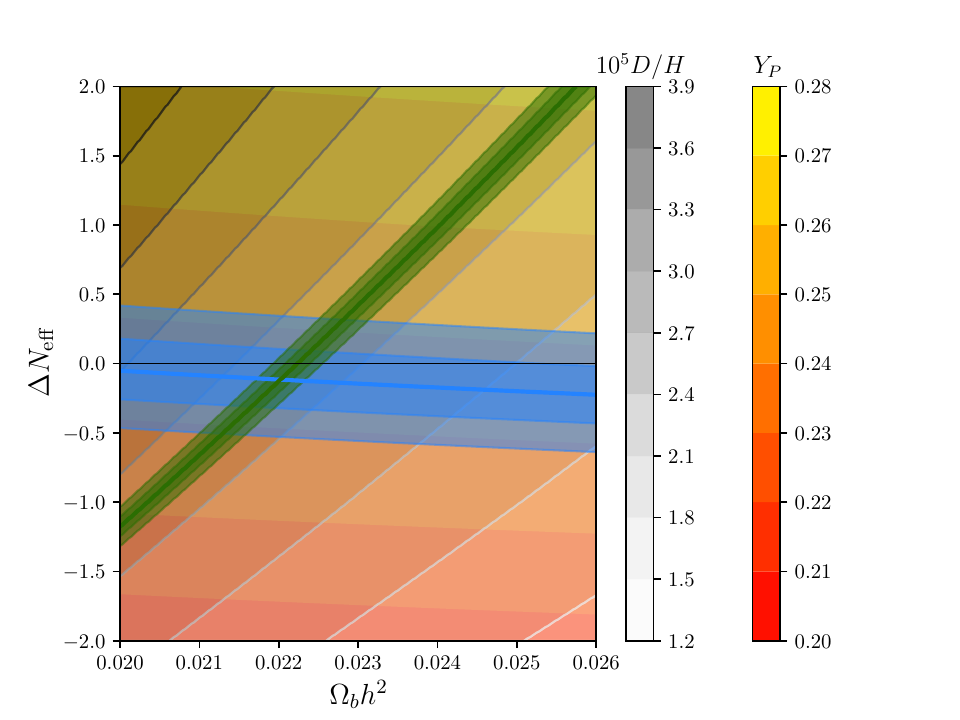}
    \caption{Light element abundances as a function of baryon abundance $\Omega_b h^2$ and number of effective additional neutrino species $\Delta N_\mathrm{eff}$. The green and blue line and contours represent the mean, and the $1\sigma$ and $2\sigma$ contours of the Deuterium and Helium recommendation from PDG \cite{ParticleDataGroup:2022pth}, respectively. Computed from the \texttt{PArthENoPE v3.0} code.}
    \label{fig:bbn_surface}
\end{figure}

\noindent These relationships are also summarized in \cref{fig:bbn_surface}. It should be noted that a higher Hubble rate will also cause a faster freeze-out of the Deuterium burning, leading to a higher Deuterium abundance. Thus the green line and the grey strips in \cref{fig:bbn_surface} are not vertical but slanted. Instead, the final Helium abundance (blue line and orange strips) does not strongly depend on the baryon abundance for the aforementioned reasons and is almost horizontal.

\section{Method and Data}\label{sec:method_data}

There are several different combinations of Deuterium and Helium measurements commonly used in the literature. In this work we make use of the data described in \cref{tab:bbn_data}, representing some of the most up-to-date measurements of light element abundances. \nils{While we cite three \enquote{separate} data sets, it needs to be stressed that there is a large degree of overlap between them, especially for \enquote{PDG Aug 2023} and \enquote{Yeh+2022}. We recommend the interested reader to look at \cref{app:data} for an overview.}

Note that we discuss the anomalous Helium measurement from \cite{Matsumoto:2022tlr} in \cref{ssec:Helium}.

\begin{table}[t]
    \centering
    \begin{tabular}{c c|c r}
        Name & Type & Value & Reference \\ \hline \rule{0pt}{1.2em}
        \multirow{2}{*}{BAO+BBN papers}& $10^5 D/H$ & $2.527 \pm 0.030$ & Cooke+2017~\cite{Cooke:2017cwo} \\
        & $Y_P$ & $0.2449 \pm 0.0040$ & Aver+2015~\cite{Aver:2015iza} \\
        \multirow{2}{*}{PDG Aug 2023}& $10^5 D/H$ & $2.547 \pm 0.029$ & PDG eq.~(24.2) \cite{ParticleDataGroup:2022pth} \\
        & $Y_P$ & $0.2450 \pm 0.0030$ & PDG eq.~(24.\nilsr{4}{3}) \cite{ParticleDataGroup:2022pth} \\
        \multirow{2}{*}{Yeh+2022~\cite{Yeh:2022heq}}& $10^5 D/H$ & $2.550 \pm 0.030$ & Yeh+2022, eq.~(1.4) ~\cite{Yeh:2022heq} \\
        & $Y_P$ & $ 0.2448 \pm 0.0033$ & Aver+2021~\cite{2022MNRAS.510..373A} \\
    \end{tabular}
    \caption{A comparison of different data sets used in the community and employed within this work. The data titled \enquote{BAO+BBN papers} was used in \cite{Schoneberg:2019wmt,Schoneberg:2022ggi}, the data titled \enquote{PDG Aug 2023} is the most recent recommendation of the particle data group (PDG) at the time of writing, and the data titled \enquote{Yeh+2022} is the one that is employed by the \cite{Yeh:2022heq}, which the PDG uses as a reference for their $\Omega_b h^2$ constraint. \nils{For notes on these data, see \cref{app:data}.}}
    \label{tab:bbn_data}
\end{table}

As far as the theoretical predictions for the light element abundances are concerned, there are two main approaches that modern codes usually adapt, related to the treatment of the underlying nuclear interaction cross sections. Either the energy/temperature-dependence of these cross sections are computed from theoretical \textit{ab-initio} computations (such as in the \texttt{PRIMAT} code \cite{Pitrou:2018cgg,Pitrou:2020etk}), or are interpolated from databases of measured values (such as in the \texttt{PArthENoPE} code \cite{Pisanti:2007hk,Consiglio:2017pot,Gariazzo:2021iiu}).\footnote{The code used in \cite{Yeh:2022heq} (which is the basis for the PDG result \cite{ParticleDataGroup:2022pth}) descends from the 1999 update \cite{Olive:1999ij} of the Wagoner code \cite{1969ApJS...18..247W}, and uses polynomial fits to the experimental data \cite{Fields:2019pfx}. There is also the code AlterBBN \cite{Arbey:2011nf,Arbey:2018zfh}, but it does not yet implement the nuclear reactions at the same level of precision. Note that the energy-dependent cross sections can be translated to temperature-dependent rates via an integral (see e.g.~\cite[eq.~(3.14)]{Serpico:2004gx} or \cite[eq.~(2)]{Yeh:2020mgl}).} For the cosmological abundances of Helium and Deuterium the most important differences between these two approaches are in the Deuterium burning reactions, namely the radiative capture ($dp\gamma$)
\begin{equation}
    \element{2}{H} + p \to \element{3}{He} + \gamma~, 
\end{equation}
and the two transfer reactions ($ddn, ddp$)
\begin{align}
    \element{2}{H} + \element{2}{H} &\to \element{3}{He} + n~, \\
    \element{2}{H} + \element{2}{H} &\to \element{3}{H} + p~, 
\end{align}
where $n$ is a neutron, $p$ is a proton, $\gamma$ a photon, $\element{2}{H}$ is Deuterium, $\element{3}{H}$ is Tritium, and $\element{3}{He}$ is Helium-3. 

Very recently the LUNA experiment has measured the $dp\gamma$ \cite{Mossa:2020gjc} radiative capture cross section at BBN energies. This allowed both methods/codes to update this crucial rate, differences remaining now mostly with the two transfer reactions $ddn$ and $ddp$ (see also \cite{Pitrou:2021vqr}). 

Even more recently, a new code has been released (\texttt{PRyMordial}~\cite{Burns:2023sgx}), which can represent both calculation types and has a very simple interface that can be used to marginalize over uncertainties related to the reaction rates.\footnote{This is on top of following neutrino decoupling and freeze-out, which allows for studies of more complicated cosmological models.} \nils{For more information on the code in general, and in particular for the implementation of and details on the marginalization, please see \cite{Burns:2023sgx}.} Note that this is also in principle possible for \texttt{PArthENoPE}, though that would require more severe modifications of the underlying likelihood code structure. 

\noindent As such, in terms of codes used to predict light element abundances, we use the following:
\begin{itemize}
    \item The old \texttt{PArthENoPE v2.0}~\cite{Consiglio:2017pot} code, which does not incorporate the radiative capture ($dp\gamma$) measurements from the LUNA experiment.
    \item The new \texttt{PArthENoPE v3.0}~\cite{2022CoPhC.27108205G} code, which does take into account these new LUNA measurements (as well as somewhat more recent measurements of $ddp,ddn$ from \cite{2014ApJ...785...96T}).
    \item The recently released \texttt{PRyMordial}~\cite{Burns:2023sgx} code, which can be run in two states:
    \begin{itemize}
        \item In its first state, it uses the NACRE II database \cite{Xu:2013fha} of thermonuclear rates for obtaining the light element abundances (updating the radiative capture with LUNA results, and the Berillium7-neutron to Lithium7-proton cross section with the fit from \cite{Fields:2019pfx})
        \item In its second state, it uses the reaction rates tabulated in the \texttt{PRIMAT} code (which themselves are based on theoretical ab-initio calculations like \cite{Descouvemont:2004cw,Iliadis:2016vkw,InestaGomez:2017eya,2019ApJ...872...75D,deSouza:2019pmr,Moscoso:2021xog}, most notably \cite{Iliadis:2016vkw} for the key Deuterium burning reactions)
    \end{itemize}
\end{itemize}
This choice of codes allows us to compare the predictions amongst all different modern BBN codes and to test the impact of the various assumptions going into the measurement. In particular, we represent the theoretical ab-initio calculations with the \texttt{PRyMordial} code in its \texttt{PRIMAT}-driven mode, and the calculations based on experimental fits will be represented either by \texttt{PArthENoPE v3.0} (without explicit marginalization over reaction rate uncertainties) or by \texttt{PRyMordial} in its NACRE II state (with explicit and more conservative marginalization over reaction rate uncertainties).\footnote{It should be noted that in all cases the normalization of the rate is fitted to experimental data, and the crucial difference between the approaches is not necessarily just the choice of shape (e.g. theoretically motivated or polynomial fit to data) but also the choice/treatment of nuclear cross section data (e.g. rejection of outliers as in \cite{InestaGomez:2017eya} used for \cite{Pitrou:2020etk} or the fitting method employed in \cite{Pisanti:2020efz}).}

\begin{figure}[h]
    \centering
    \includegraphics[width=0.75\textwidth]{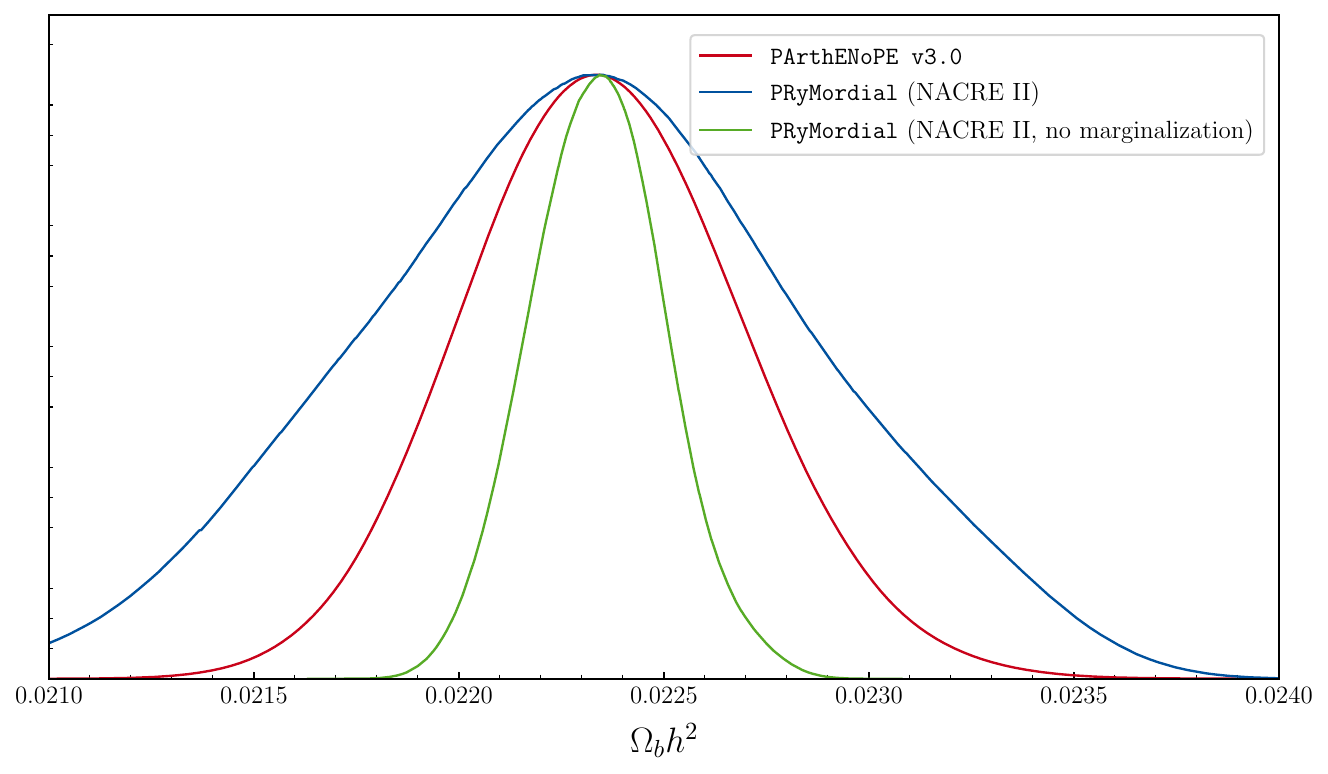}
    \caption{Impact of marginalization over nuclear reaction rates for the \texttt{PRyMordial} code, compared to the \texttt{PArthENoPE v3.0} adopted uncertainties (as described in the text). The remaining width of the \texttt{PRyMordial} code without marginalization directly represents the uncertainty of the deuterium determination.}
    \label{fig:bbn_marg}
\end{figure}
It should be noted that in \cite{Schoneberg:2019wmt,Schoneberg:2022ggi} and here for \texttt{PArthENoPE} there is no explicit marginalization over the uncertainties over the nuclear rates. Instead, an approximate theoretical uncertainty on the final abundance was derived from the main results reported in \cite{Gariazzo:2021iiu} ($6\cdot 10^{-7}$ for $D/H$). In the results derived from the \texttt{PRyMordial} code the marginalization is instead taken into account explicitly, by marginalizing over log-normally distributed rate uncertainties as described in \cite{Burns:2023sgx}. If this marginalization was not performed, the uncertainty would be dramatically smaller (dominated instead only by the experimental uncertainty), as we show in \cref{fig:bbn_marg}. It should be also noted that this marginalization performed in the \texttt{PRyMordial} code is overall more conservative compared to the rate uncertainties adopted in \cite{Gariazzo:2021iiu}, which are derived from those discussed in \cite{Pisanti:2020efz}. \nils{This very conservative treatment is useful if a baryon abundance value is required for cosmological inference, but one wants to remain agnostic with respect to the well known and well documented systematic differences arising from different treatment of the nuclear rates (discussed above).}

We put flat priors on the baryon abundance and the additional neutrino number (where applicable). The neutron lifetime is varied freely between 876.4s and 882.4s with a flat prior, conservatively encompassing the currently measured value from bottle experiments ($877.5^{+0.5}_{-0.44}$s \cite{UCNt:2021pcg}).\footnote{In any case we do not find a strong correlation of the neutron lifetime with the final derived baryon abundance.  Even when the lifetime is fixed to 879.4s (PDG 2019 \cite[2019 update]{PhysRevD.98.030001}), the baryon abundance is very similar. A degeneracy is expected if much larger variations of the lifetime of around $\mathcal{O}(100\mathrm{s})$ would be taken into account (see for example \cite[Fig.~1]{Chowdhury:2022ahn})}

\section{Results}\label{sec:results}

In this section we find the results for various combinations of codes and data, in order to find conservative and reliable estimates of the current baryon abundance (and the effective number of neutrinos). For this, we investigate in \cref{ssec:codes} the various codes and in \cref{ssec:data} the different available datasets, discussing the anomalous Helium measurement in \cref{ssec:Helium}. Finally, we discuss the impact of varying the effective number of neutrinos in \cref{ssec:neutrinos}. A summary of all results may be found in \cref{tab:results}.

\subsection{Sensitivity to employed code}\label{ssec:codes}

\begin{figure}[h]
    \centering
    \includegraphics[width=0.65\textwidth]{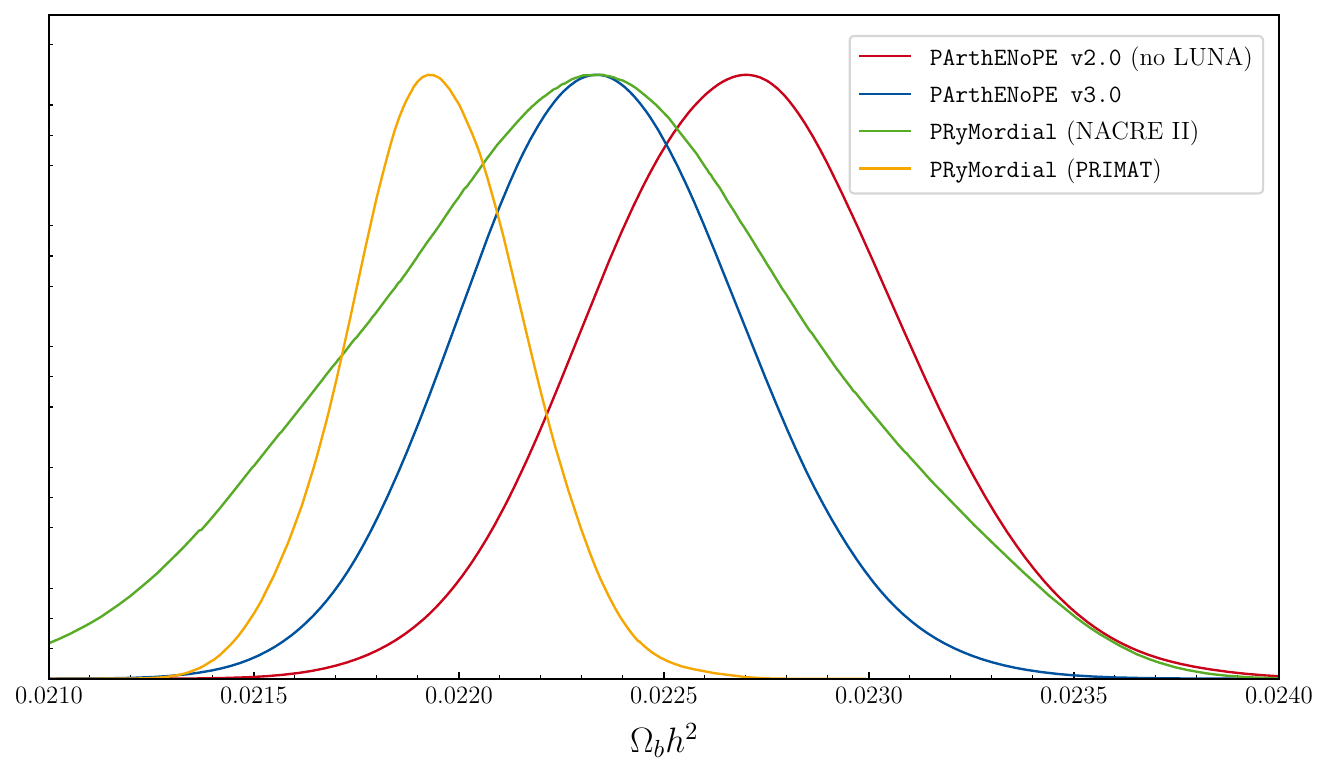}
    \caption{Comparison of the $\Omega_b h^2$ constraints in the $\Lambda$CDM model between different codes (or settings) for the same underlying data titled \enquote{BAO+BBN papers} in \cref{tab:bbn_data}.}
    \label{fig:bbn_codes}
\end{figure}

As a first step we compare the old results from the \texttt{PArthENoPE v2.0} and \texttt{PArthENoPE v3.0} codes to highlight the impact of the deuterium radiative capture rate ($dp\gamma$) as measured by LUNA \cite{Mossa:2020gjc}. For this purpose we use the data from the BAO+BBN papers (see \cref{tab:bbn_data}). This comparison is shown in \cref{fig:bbn_codes}, comparing the red versus blue lines. 

It is immediately evident that there is a large impact of the new measurements of the nuclear rate --- the different deuterium burning rates lead to a decreased deuterium abundance for a given fixed $\Omega_b h^2$, requiring lower baryon abundances to produce the same deuterium abundance due to their anti-correlation --- which had also been pointed out in \cite{Schoneberg:2022ggi}. The result is a shift in the mean of $\Omega_b h^2$ by around $-0.9\sigma$ from $\Omega_b h^2 = 0.02271\pm 0.00038$ with the old code to $\Omega_b h^2 = 0.02236 \pm 0.00034$ with the new code without a significant change in the uncertainty (see \cite{Schoneberg:2022ggi} for a discussion on the uncertainties).\footnote{These numbers differ marginally from those reported in \cite{Schoneberg:2022ggi} due to the method used to derive them from the MCMC chains. In \cite{Schoneberg:2022ggi} the maximum-posterior intervals as derived by \texttt{MontePython} \cite{Brinckmann:2018cvx} are reported, whereas here we report the mean and square root of the variance for the given chains. Additionally, for the \texttt{PArthENoPE v3.0} run, new interpolation tables have been derived including also the neutron lifetime ($\tau_n$).}

\noindent There are two important questions to answer at this point:
\begin{enumerate}
    \item Is there a large difference between the different nuclear rate computations? This question is answered by comparing the blue and yellow lines in \cref{fig:bbn_codes} (comparing \texttt{PArthENoPE v3.0} as a stand-in for the experimental rates and \texttt{PRyMordial} in its \texttt{PRIMAT}-driven mode as a stand-in for the theoretical rates). One can observe that there is indeed a decent shift, from $\Omega_b h^2 = 0.02236 \pm 0.00034$ to $\Omega_b h^2 = 0.02195 \pm 0.00021$,\footnote{This almost exactly reproduces the published \texttt{PRIMAT} results of \cite{Pitrou:2020etk} and the published \texttt{PArthENoPE} results of \cite{Gariazzo:2021iiu}.} the two results differing in mean by around $1.0\sigma$ (and the uncertainty being smaller by around 40\%).
    \enlargethispage*{2\baselineskip}
    \item Can the conservative marginalization over larger experimental uncertainties bring the two results back into agreement? This question is answered by comparing the blue, green, and yellow lines in \cref{fig:bbn_codes} (the green line represents the \texttt{PRyMordial} code in its experimentally driven mode with the additional marginalization over the uncertainties). Indeed, we find that the corresponding constraint of $\Omega_b h^2 = 0.02231 \pm 0.00055$ covers all of these results nicely. The corresponding shifts are $+0.08\sigma$ for \texttt{PArthENoPE v3.0} and $-0.6\sigma$ for \texttt{PRyMordial} in its \texttt{PRIMAT}-driven mode (and $+0.6\sigma$ for \texttt{PArthENoPE v2.0}).
\end{enumerate}
To summarize, without marginalization over experimental uncertainties there is a somewhat significant difference between the theoretical \textit{ab-initio} calculations and the experimental calculations, but when the marginalization feature of \texttt{PRyMordial} is used, the results largely agree between the codes.

\subsection{Sensitivity to employed datasets}\label{ssec:data}

\begin{figure}[H]
    \centering
    \includegraphics[width=0.45\textwidth]{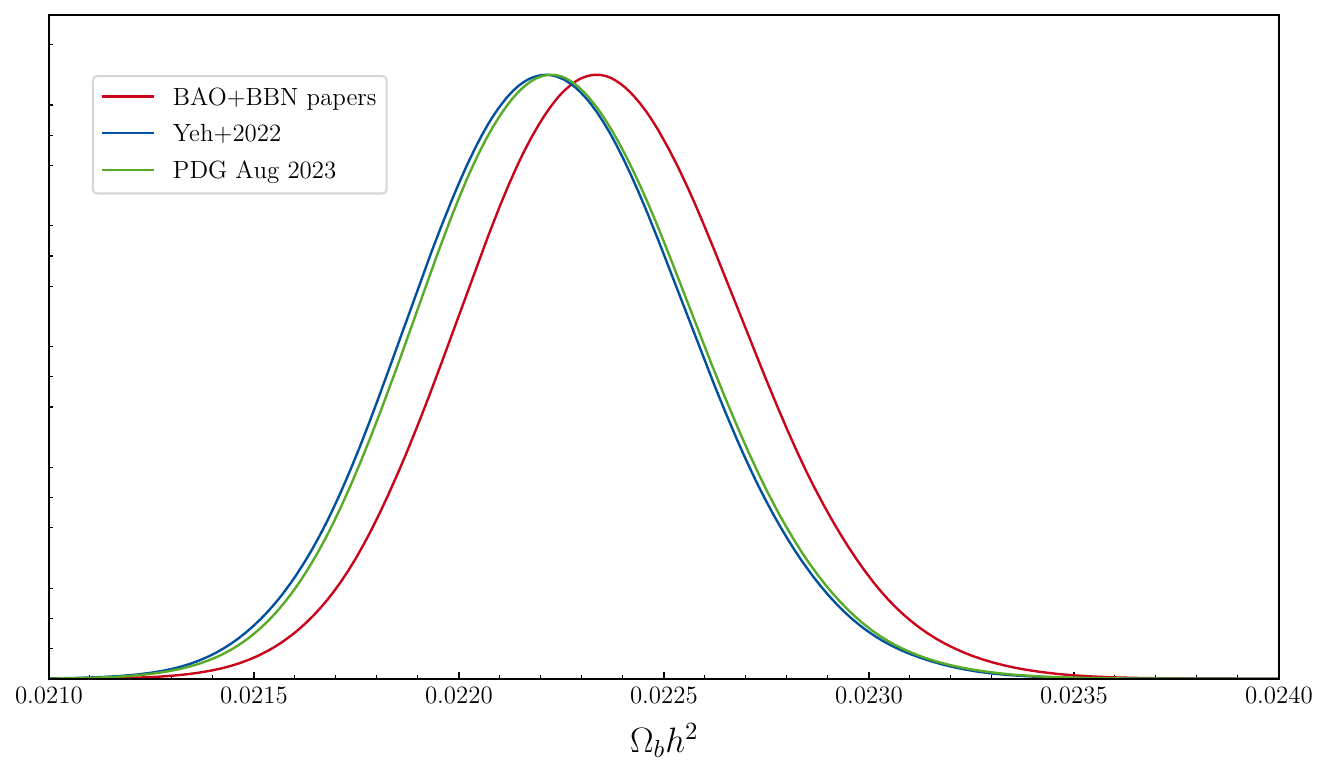} 
    \includegraphics[width=0.45\textwidth]{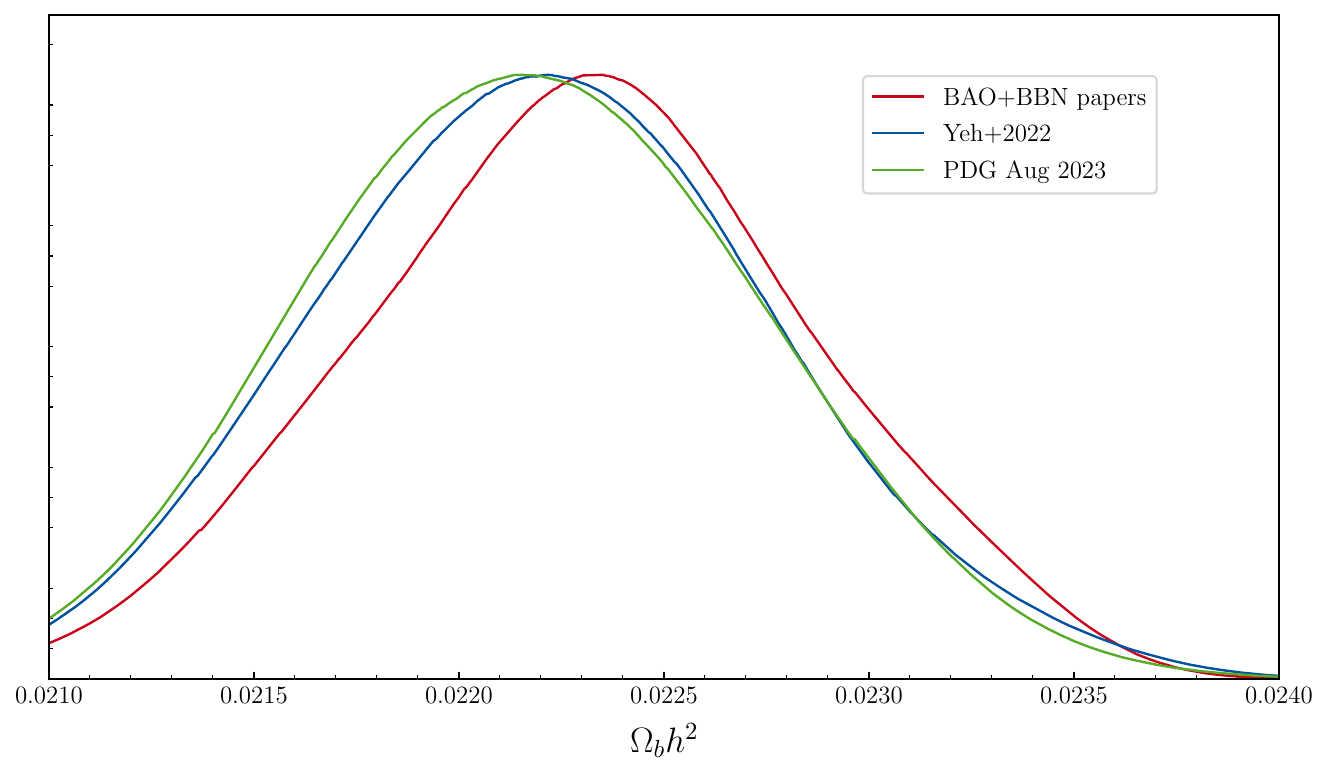} \\
    \includegraphics[width=0.45\textwidth]{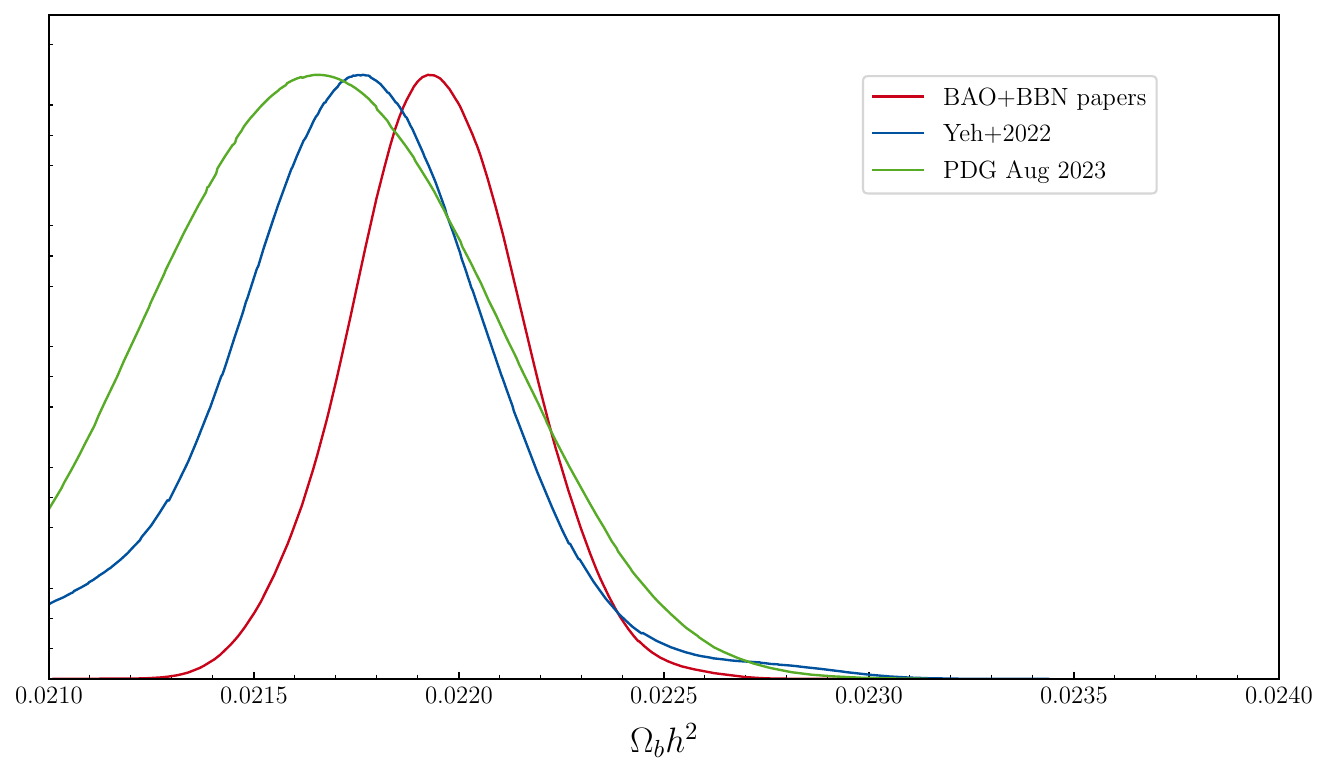}
    \caption{Comparison of the baryon abundance constraints from different data sets when using the \texttt{PArthENoPE v3.0} code (top left), using the \texttt{PRyMordial} code in NACRE II mode (top right), and using the \texttt{PRyMordial} code in its \texttt{PRIMAT} mode (bottom). For the datasets, see \cref{tab:bbn_data}.}
    \label{fig:bbn_datas}
\end{figure}

The second kind of difference to investigate is that based on the adopted Deuterium (and Helium) abundance data. In \cref{fig:bbn_datas} we show the baryon abundances derived for the different datasets of \cref{tab:bbn_data} for each of the three adopted codes. The corresponding baryon abundance shifts only very slightly for the different light element datasets, typically by less than $0.3\sigma$. 

However, the abundance of the \texttt{PRyMordial} code used in its \texttt{PRIMAT} mode does shift more drastically. There we do see a shift up to $1.4\sigma$ towards lower values of baryon abundance when different light element abundances are used.

\enlargethispage*{6\baselineskip}

\subsection{The Helium anomaly}\label{ssec:Helium}

Recently the Helium abundance has been measured by the EMPRESS survey performed on the Subaru telescope \cite{Matsumoto:2022tlr}, giving results that are quite a bit lower ($Y_P = 0.2370 \pm 0.0034$, at $-1.8\sigma$ compared to the PDG recommended value \cite{ParticleDataGroup:2022pth}) than any other recent measurement in the literature \cite{ParticleDataGroup:2022pth}. It is thus important to clarify if a much lower Helium abundance has any drastic impact on the derived baryon abundance. 

Depending on whether or not marginalization over the nuclear rates is taken into account, the results are somewhat different. 
To gain a better understanding, we first look at the case without marginalization, where the shift is rather minor ($-0.1\sigma$ for \texttt{PArthENoPE v3.0}). However, there is an incompatibility of the data and the model which becomes evident in the maximum likelihood. For example, while the minimum $\chi^2 = - 2 \ln \mathcal{L}$ reached for the run assuming the PDG-recommended abundances is 0 to the numerical precision of around $\Delta \chi^2 \sim 0.1$ of the MCMC chains, it is around $\chi^2_\mathrm{min} \simeq 4.6$ if the EMPRESS Helium abundance is forced.

\begin{figure}[t]
    \centering
    \includegraphics[width=0.65\textwidth]{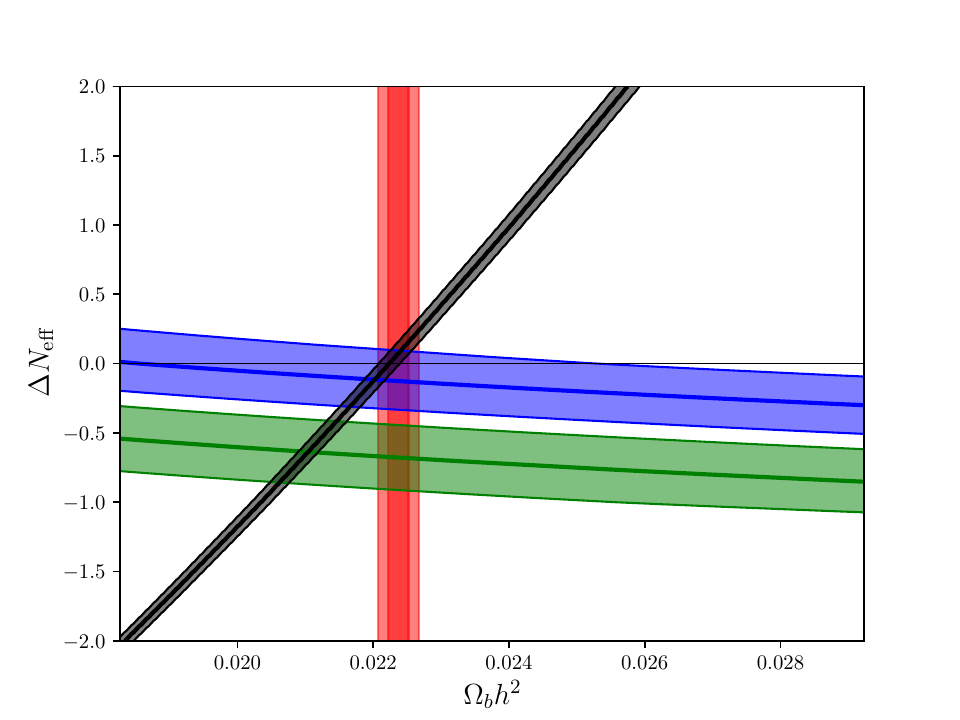}
    \caption{A comparison of various constraints in the $\Delta N_\mathrm{eff}-\Omega_b h^2$ plane. In black and blue we show the parameters compatible with the PDG-recommended measurements for the Deuterium and Helium abundances (mean+$1\sigma$), respectively (see \cref{tab:bbn_data}). In green we show the parameters compatible with the EMPRESS Helium abundance measurement (mean+$1\sigma$). The BBN computations were performed here with \texttt{PArthENoPE v3.0}. Finally, the red contours show the baryon abundance ($1\sigma$ and $2\sigma$) from Planck \cite{Planck:2018vyg} as a comparison.}
    \label{fig:bbn_helium}
\end{figure}
This incompatibility is most easily visualized when extending the parameter space to include additional free-streaming neutrinos through the $\Delta N_\mathrm{eff}$ parameter as in \cref{fig:bbn_helium}. The only intersection of the Helium abundance and the Deuterium abundance (black contour) that is compatible with the assumption of no additional dark neutrinos (the thin black line at $\Delta N_\mathrm{eff}=0$) is for the PDG recommended value of the Helium abundance (blue contour). Instead, the EMPRESS-derived value of the Helium abundance is not compatible with the measured Deuterium abundance and simultaneously $\Delta N_\mathrm{eff} = 0$ in this extended parameter space.

Instead, if the \texttt{PRyMordial} code is used in either of its modes, the shift is around $-1.2\sigma$ (see \cref{tab:results}). This is because the broad simultaneous marginalization over nuclear rates and the neutron lifetime can put such low Helium abundances back into accordance at the cost of putting the various rates at the edges of their experimentally/theoretically allowed regions. If instead we fix the neutron lifetime to a central value of 879.4s in these cases, for example, the results are again within $0.3\sigma$ to their baseline values.\enlargethispage*{3\baselineskip}

As such, in this work we will treat the Helium abundance measurement of \cite{Matsumoto:2022tlr} as an 'outlier' until further evidence emerges.

\subsection{Additional relativistic degrees of freedom}\label{ssec:neutrinos}

\begin{figure}[t]
    \centering
    \includegraphics[width=0.3\textwidth]{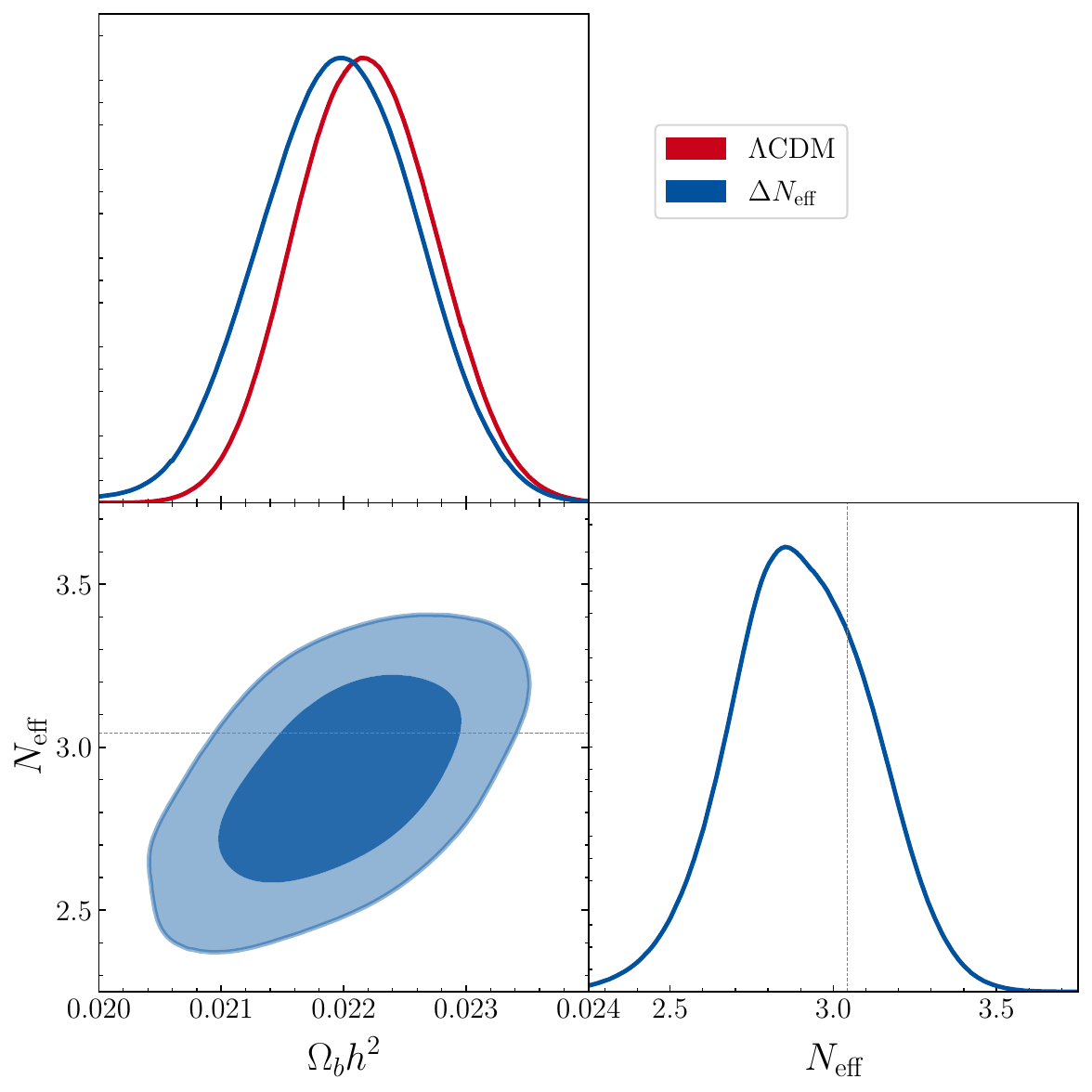}
    \includegraphics[width=0.3\textwidth]{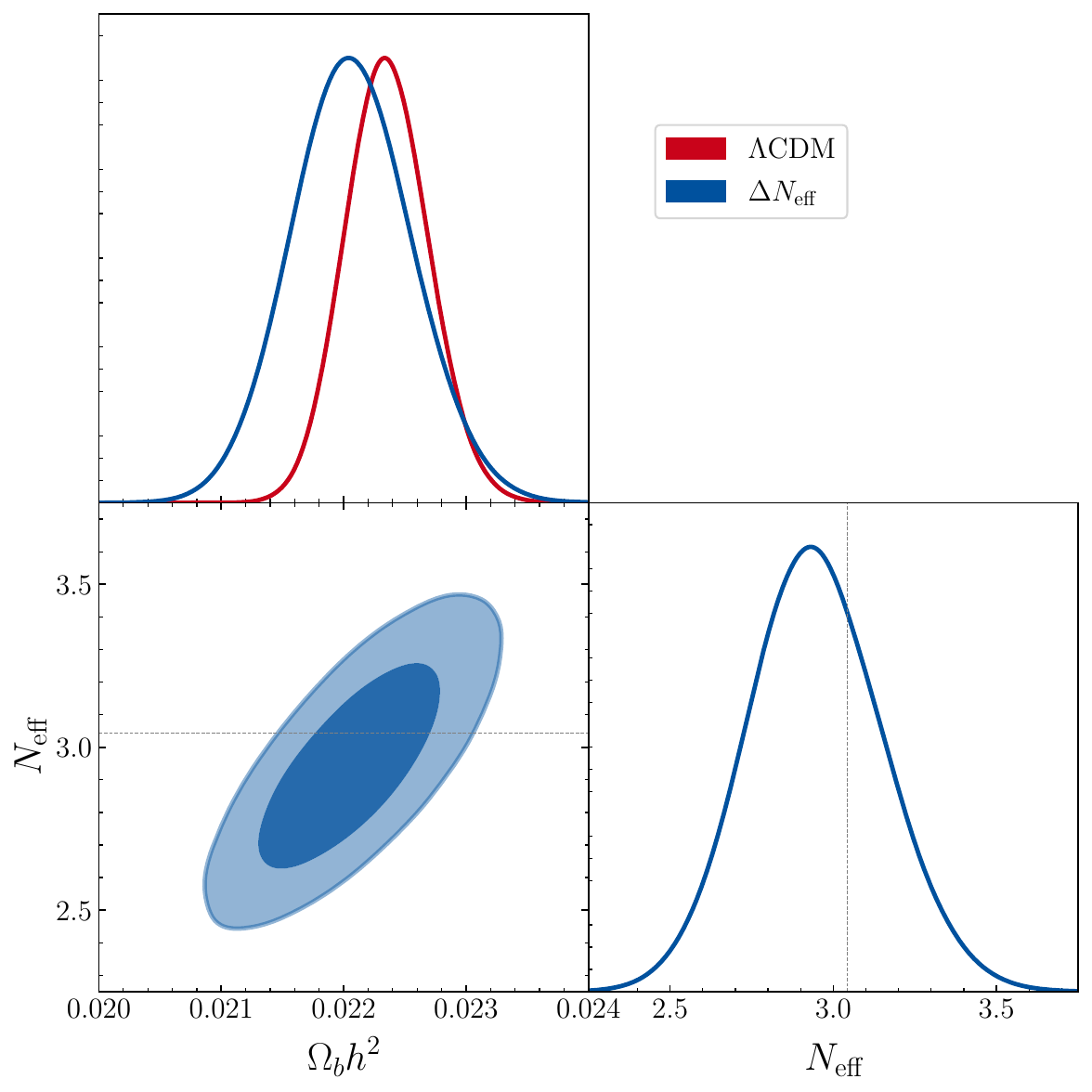}
    \includegraphics[width=0.3\textwidth]{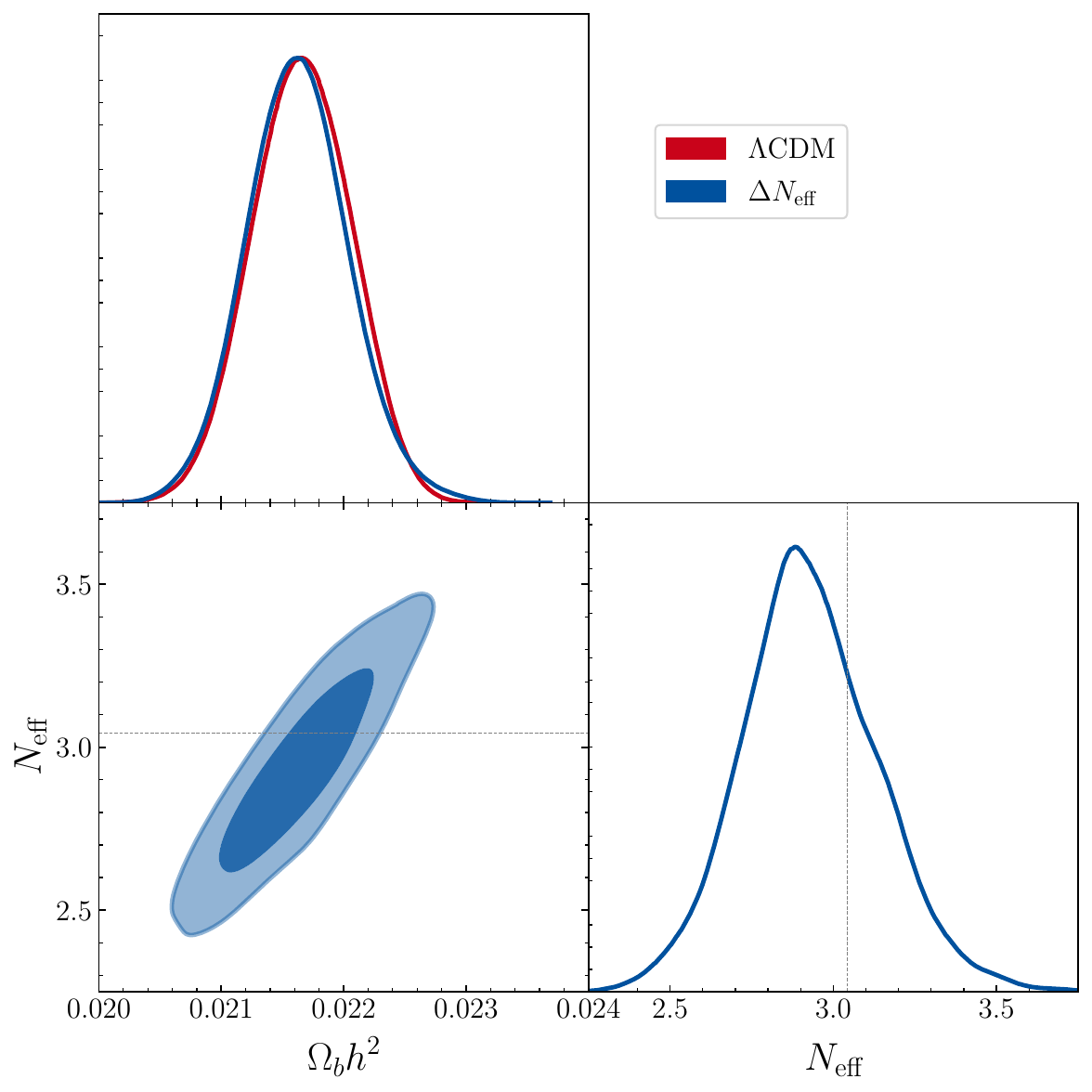}
    \caption{Constraints from the various codes for the joint $N_\mathrm{eff}-\Omega_b h^2$ parameter space. \textbf{Left.} \texttt{PRyMordial} (NACRE II). 
 \textbf{Middle.} \texttt{PArthENoPE v3.0}. \textbf{Right.} \texttt{PRyMordial} (PRIMAT).}
    \label{fig:bbn_neutrino}
\end{figure}

The inclusion of additional neutrino-like species through the parameter $\Delta N_\mathrm{eff}$ has a large impact on BBN and can correspondingly be constrained by the light element abundances, providing a constraint of similar strength as from the CMB. However, this constraint is instead much more dependent on the assumed Helium abundance, as visible in \cref{fig:bbn_helium} and as described in \cref{sec:theory}. In \cref{fig:bbn_neutrino} we show the two-dimensional constraints obtained from the various codes for the PDG recommended abundances from \cref{tab:bbn_data}. It is immediately evident that the inclusion of additional relativistic degrees of freedom during BBN does not significantly bias the constraints on the baryon abundance, typically leading to a mild widening of the contours as well as a slight downward shift of the mean value (see also \cref{tab:results}). 

The constraint on $\Omega_b h^2$ degrades to around $\Omega_b h^2 = 0.02196 \pm 0.00063$ with PDG-recommended light element abundances once marginalizing over $\Delta N_\mathrm{eff}$\,, though a more conservative estimate is given by the older Helium abundance measurements from \cite{Aver:2015iza} that were used in the BAO+BBN papers, leading to $\Omega_b h^2 = 0.02212 \pm 0.00072$. In any case, almost all variations are well within one sigma of each other, and there is no obvious bias.

The constraint on the $\Delta N_\mathrm{eff}$ parameter depends on the tightness of the assumed Helium abundance (as expected from \cref{sec:theory}), with the constraints ranging from $-0.10 \pm 0.21$ (for PDG-recommended abundances and the \texttt{PArthENoPE v3.0} code) up to $-0.09 \pm 0.28$ (for the abundances used in the BAO+BBN papers and the \texttt{PArthENoPE v3.0} code). See \cref{tab:results} for further results.

\begin{table}[t]
    \centering
    \begin{tabular}{l l | c c}
         Data & Code & $100 \Omega_b h^2$ & $\Delta N_\mathrm{eff}$ \\ \hline\newtabline
         \multirow{5}{*}{BAO+BBN papers}
         & \texttt{PArthENoPE v3.0} & $2.236 \pm 0.034$ & --- \\
         & \texttt{PRyMordial} (NACRE II) & $2.231 \pm 0.055$ & --- \\
         & \texttt{PRyMordial} (PRIMAT) &  $2.195 \pm 0.021$ & --- \\ \cline{2-4}
         &\texttt{PArthENoPE v2.0} & $2.271\pm 0.038$ & --- \\
         & \texttt{PRyMordial} (NACRE II, no marginalization) & $2.234 \pm 0.016$ & --- \\ \newtabline
         \multirow{3}{*}{PDG} 
          & \texttt{PArthENoPE v3.0} & $2.225 \pm 0.033$ & --- \\
           & \texttt{PRyMordial} (NACRE II)  & $2.218 \pm 0.055 $& --- \\
         & \texttt{PRyMordial} (PRIMAT) &  $2.166 \pm 0.039$ & --- \\ \newtabline
         \multirow{3}{*}{Yeh+2022} 
         & \texttt{PArthENoPE v3.0} & $2.223 \pm 0.034$ & --- \\
          &\texttt{PRyMordial} (NACRE II)  & $2.221 \pm 0.056$ & --- \\
         & \texttt{PRyMordial} (PRIMAT) &  $2.173 \pm 0.036$ & --- \\ \hline\newtabline
         \multirow{3}{*}{BAO+BBN papers} 
          & \texttt{PArthENoPE v3.0 ($+\Delta N_\mathrm{eff}$)} & $2.218 \pm 0.064$ & $-0.09\pm 0.28$ \\
           & \texttt{PRyMordial} (NACRE II, +$\Delta N_\mathrm{eff}$)  & $2.212 \pm 0.072$ & $-0.12 \pm 0.27$\\
         & \texttt{PRyMordial} (PRIMAT, +$\Delta N_\mathrm{eff}$) & $2.172 \pm 0.055$  &  $-0.13\pm0.28$\\ \newtabline
         \multirow{3}{*}{PDG} 
          &  \texttt{PArthENoPE v3.0 ($+\Delta N_\mathrm{eff}$)} & $2.207 \pm 0.050$ & $-0.10 \pm 0.21$ \\
           & \texttt{PRyMordial} (NACRE II, +$\Delta N_\mathrm{eff}$)  & $2.196 \pm 0.063$ & $-0.14 \pm 0.21$ \\
         & \texttt{PRyMordial} (PRIMAT, +$\Delta N_\mathrm{eff}$) & $2.163 \pm 0.042$   & $-0.11 \pm 0.21$ \\ \newtabline
         \multirow{3}{*}{Yeh+2022} 
          & \texttt{PArthENoPE v3.0 ($+\Delta N_\mathrm{eff}$)} & $2.203 \pm 0.053$ & $-0.11\pm0.23$ \\
           & \texttt{PRyMordial} (NACRE II, +$\Delta N_\mathrm{eff}$)  & $2.194 \pm 0.064$& $-0.13 \pm 0.23$ \\
         & \texttt{PRyMordial} (PRIMAT, +$\Delta N_\mathrm{eff}$) & $2.157 \pm 0.048$ & $-0.14\pm0.24$ \\ \hline \newtabline
          \rule{0pt}{2em} 
         \multirow{6}{*}{EMPRESS} 
         & \texttt{PArthENoPE v3.0} & $ 2.221 \pm 0.033$ & --- \\& \texttt{PRyMordial} (NACRE II) & $2.128 \pm 0.048$& --- \\
         & \texttt{PRyMordial} (NACRE II, fixed neutron lifetime) & $2.194 \pm 0.071$& --- \\
         & \texttt{PRyMordial} (NACRE II, no marginalization) & $2.168 \pm 0.053$& --- \\
          & \texttt{PRyMordial} (PRIMAT) & $2.100 \pm 0.042$& --- \\
           & \texttt{PRyMordial} (PRIMAT, no marginalization) & $2.177 \pm 0.024$& ---\\
    \end{tabular}
    \caption{The numerical value of the constraints (mean and standard deviation) of the various runs considered in this paper.}
    \label{tab:results}
\end{table}
\section{Conclusions}\label{sec:conclusions}

The state of the baryon abundance as inferred from measurements of the light element abundances from big bang nucleosynthesis is at this point relatively clear. While there is certainly still a spread of the constraints between various codes (and their underlying assumptions about the Deuterium burning rates), the field has come a long way of more systematically characterizing these uncertainties and biases. With relatively conservative assumptions about the reaction rate uncertainties it is possible to derive a baryon abundance estimate that nicely covers all available results in terms of available codes and datasets. For this purpose, we recommend the value derived by the \texttt{PRyMordial} code using the NACRE II database, giving
\begin{equation}
 \Omega_b h^2 = 0.02218 \pm 0.00055~,
\end{equation}
with PDG-recommended abundances \cite{ParticleDataGroup:2022pth}, and
\begin{equation}
    \Omega_b h^2 = 0.02231 \pm 0.00055~,
\end{equation}
with the most recent Deuterium determination from \cite{Cooke:2017cwo}. 

Notably, the inflated uncertainties compared to the analyses performed in \cite{ParticleDataGroup:2022pth,Yeh:2022heq,Pitrou:2020etk} and many others is related to a broader marginalization over the uncertainties in the nuclear rates, leading to more conservative estimates that encompass both the theoretical ab-initio calculations and the experimental rates for the $ddn$ and $ddp$ Deuterium burning processes.

These values are largely insensitive to the assumed Helium abundance, as long as it remains within a range compatible with the standard BBN codes. The EMPRESS result \cite{Matsumoto:2022tlr} would correspond to a much lower abundance, but this is an effect of the breakdown of the compatibility with standard rates and assumptions of the BBN codes -- which can be circumvented only in more exotic neutrino models (see for example \cite{Burns:2022hkq,Takahashi:2022cpn}). This result is also (mildly) incompatible with the collection of all previous results. As such, until further evidence in this direction emerges, we consider this result an 'outlier'.

As far as more exotic cosmologies with additional neutrinos are concerned, we find that here too results are mostly consistent, pointing to roughly the same baryon abundances. We find in $\Lambda$CDM+$\Delta N_\mathrm{eff}$ that
\begin{equation}
    \Omega_b h^2 = 0.02196 \pm 0.00063
\end{equation} with PDG-recommended abundances, and 
\enlargethispage*{3\baselineskip}
\begin{equation}
    \Omega_b h^2 = 0.02212 \pm 0.00072
\end{equation} with the older estimates from the BAO+BBN papers) and the amount of additional neutrino species constrained at the level of around $\Delta N_\mathrm{eff} = -0.1 \pm 0.2$ (see \cref{tab:results} for precise values).

While further cosmological measurements of the Deuterium and Helium abundances will also continue to be necessary (especially to cross-check the previous measurements), the largest advances in our understanding of the light element abundances from big bang nucleosynthesis are expected to come from further laboratory measurements of the Deuterium burning rates, which currently represent the most crucial uncertainty in the estimation of the baryon abundance. As such, we expect the future of the baryon abundance as determined from big bang nucleosynthesis to bring even higher accuracy and precision.

\acknowledgments{The author very warmly thanks Prof. L. Verde for her ideas, suggestions and support, without which this work would not have been possible. The author also thanks Prof. J. Lesgourgues, Prof. C. Pitrou, Profs. M. Valli, 
A. Burns, T.M.P. Tait, and Profs. O. Pisanti and G. Mangano for their feedback on the draft. NS acknowledges the support of the following Maria de Maetzu fellowship grant: Esta publicaci\'on es parte de la ayuda CEX2019-000918-M, financiada por MCIN/AEI/10.13039/501100011033.}

\appendix
\section{Notes on the employed data sets}\label{app:data}

\nils{
First and foremost, it needs to be stressed that there is a large overlap between the data sets, and hence the results from variations of data sets cannot be seen as independent.

First, $Y_P$ is commonly measured in extragalactic HII regions. We note that the $Y_P$ data of Aver+2021~\cite{2022MNRAS.510..373A} used for \enquote{Yeh+2022} includes the data of Aver+2015~\cite{Aver:2015iza} used for \enquote{BAO+BBN papers}, additionally incorporating LeoP (already in Aver+2020~\cite{Aver:2020fon}) and AGC 198691. The \enquote{PDG Aug 2023} case incorporates the results of Aver+2020~\cite{Aver:2020fon}, and additionally those of \cite{Valerdi:2019beb,Fernandez:2019hds,2020ApJ...896...77H,Kurichin:2021ppm,2021MNRAS.505.3624V}, though taking into account their partial overlap. It is striking to see that the results are so similar, despite the large increase of investigated systems (see \cite{ParticleDataGroup:2022pth}). To see the effect of changing this value drastically, see \cref{ssec:Helium} and the corresponding entries in \cref{tab:results}.

Second, $D_H$ is commonly measured from damped Lyman-$\alpha$ systems. 
The $D_H$ data from \enquote{PDG Aug 2023} include all of the  16 distinct measurements in \cite{Pettini:2001yu,DOdorico:2001vgo,Srianand:2010un,Fumagalli:2011iw,Noterdaeme:2012pa,Cooke:2013cba,Riemer-Sorensen:2014aoa,Balashev:2015hoe,Riemer-Sorensen:2017pey,Cooke:2017cwo,Zavarygin:2017cov}, but only uses the 11 most recent of these for their $D_H$ recommendation. Instead, the $D_H$ data from \enquote{Yeh+2022} cite only the subset of references \cite{Cooke:2013cba,Riemer-Sorensen:2014aoa,Balashev:2015hoe,Riemer-Sorensen:2017pey,Cooke:2017cwo} for their value, but also add \cite{2012MNRAS.425.2477P,2018MNRAS.477.5536Z,Cooke:2016rky}. However, the given result of their eq.~(1.4) is neither the weighted nor unweighted average of these measurements. One possible explanation would be obtained by rounding the measurement of \cite{Cooke:2016rky} (summarizing 6 systems, leading to $D_H = 2.547\pm0.033$) to the nearest digit of precision, but the authors of \cite{Yeh:2022heq} do not describe their procedure of summarizing the $D/H$ data in more detail, making a more detailed comparison difficult. Finally, the \enquote{BAO+BBN papers} data set includes only \cite{Cooke:2017cwo}, which itself summarizes 7 systems (the 6 of \cite{Cooke:2016rky}, and one additional). Note that the difference of \cite{Cooke:2016rky} and \cite{Cooke:2017cwo} at the level of $10^5 \Delta (D/H) = -0.02$ (around $0.7\sigma$) is caused by the one additional system (Q1243+307) under investigation there. As obvious from \cite[Tab.~24.1]{ParticleDataGroup:2022pth} there is a great consistency of the measured D/H value in all systems measured less than around 2013. In particular, there are only three systems system further than $1.1\sigma$ away from the mean value of the PDG recommended value; QSO J1444+2919 measured by \cite{Balashev:2015hoe} (deviating by $-1.7\sigma$), QSO 1243+307 by \cite{Cooke:2017cwo} (deviating by $-1.57\sigma$), and PKS 1937-1009 by \cite{Riemer-Sorensen:2017pey} (deviating by $+1.4\sigma$). Of course it is entirely to be expected that out of 11 measurements 3 can be beyond their common mean by more than $1\sigma$ deviation, allowing us to conclude that the D/H measurements are also remarkably consistent.
}
\bibliography{biblio}

\providecommand{\href}[2]{#2}\begingroup\raggedright\begin{thebibliography}{10}

\bibitem{Planck:2018vyg}
{\scshape Planck} collaboration, \emph{{Planck 2018 results. VI. Cosmological
  parameters}},
  \href{https://doi.org/10.1051/0004-6361/201833910}{\emph{Astron. Astrophys.}
  {\bfseries 641} (2020) A6}
  [\href{https://arxiv.org/abs/1807.06209}{{\ttfamily 1807.06209}}].

\bibitem{Riess:2021jrx}
A.G.~Riess et~al., \emph{{A Comprehensive Measurement of the Local Value of the
  Hubble Constant with 1 km s${}^{-1}$ Mpc${}^{-1}$ Uncertainty from the Hubble
  Space Telescope and the SH0ES Team}},
  \href{https://doi.org/10.3847/2041-8213/ac5c5b}{\emph{Astrophys. J. Lett.}
  {\bfseries 934} (2022) L7}
  [\href{https://arxiv.org/abs/2112.04510}{{\ttfamily 2112.04510}}].

\bibitem{Verde:2023lmm}
L.~Verde, N.~Sch\"oneberg and H.~Gil-Mar\'\i{}n, \emph{{A tale of many $H_0$}},
   \href{https://arxiv.org/abs/2311.13305}{{\ttfamily 2311.13305}}.

\bibitem{Freedman:2023jcz}
W.L.~Freedman and B.F.~Madore, \emph{{Progress in direct measurements of the
  Hubble constant}},
  \href{https://doi.org/10.1088/1475-7516/2023/11/050}{\emph{JCAP} {\bfseries
  11} (2023) 050} [\href{https://arxiv.org/abs/2309.05618}{{\ttfamily
  2309.05618}}].

\bibitem{Pisanti:2020efz}
O.~Pisanti, G.~Mangano, G.~Miele and P.~Mazzella, \emph{{Primordial Deuterium
  after LUNA: concordances and error budget}},
  \href{https://doi.org/10.1088/1475-7516/2021/04/020}{\emph{JCAP} {\bfseries
  04} (2021) 020} [\href{https://arxiv.org/abs/2011.11537}{{\ttfamily
  2011.11537}}].

\bibitem{Yeh:2020mgl}
T.-H.~Yeh, K.A.~Olive and B.D.~Fields, \emph{{The impact of new $d(p,\gamma)$3
  rates on Big Bang Nucleosynthesis}},
  \href{https://doi.org/10.1088/1475-7516/2021/03/046}{\emph{JCAP} {\bfseries
  03} (2021) 046} [\href{https://arxiv.org/abs/2011.13874}{{\ttfamily
  2011.13874}}].

\bibitem{Pitrou:2020etk}
C.~Pitrou, A.~Coc, J.-P.~Uzan and E.~Vangioni, \emph{{A new tension in the
  cosmological model from primordial deuterium?}},
  \href{https://doi.org/10.1093/mnras/stab135}{\emph{Mon. Not. Roy. Astron.
  Soc.} {\bfseries 502} (2021) 2474}
  [\href{https://arxiv.org/abs/2011.11320}{{\ttfamily 2011.11320}}].

\bibitem{Addison:2013haa}
G.E.~Addison, G.~Hinshaw and M.~Halpern, \emph{{Cosmological constraints from
  baryon acoustic oscillations and clustering of large-scale structure}},
  \href{https://doi.org/10.1093/mnras/stt1687}{\emph{Mon. Not. Roy. Astron.
  Soc.} {\bfseries 436} (2013) 1674}
  [\href{https://arxiv.org/abs/1304.6984}{{\ttfamily 1304.6984}}].

\bibitem{Aubourg:2014yra}
E.~Aubourg et~al., \emph{{Cosmological implications of baryon acoustic
  oscillation measurements}},
  \href{https://doi.org/10.1103/PhysRevD.92.123516}{\emph{Phys. Rev. D}
  {\bfseries 92} (2015) 123516}
  [\href{https://arxiv.org/abs/1411.1074}{{\ttfamily 1411.1074}}].

\bibitem{Addison:2017fdm}
G.E.~Addison, D.J.~Watts, C.L.~Bennett, M.~Halpern, G.~Hinshaw and
  J.L.~Weiland, \emph{{Elucidating $\Lambda$CDM: Impact of Baryon Acoustic
  Oscillation Measurements on the Hubble Constant Discrepancy}},
  \href{https://doi.org/10.3847/1538-4357/aaa1ed}{\emph{Astrophys. J.}
  {\bfseries 853} (2018) 119}
  [\href{https://arxiv.org/abs/1707.06547}{{\ttfamily 1707.06547}}].

\bibitem{Blomqvist:2019rah}
M.~Blomqvist et~al., \emph{{Baryon acoustic oscillations from the
  cross-correlation of Ly$\alpha$ absorption and quasars in eBOSS DR14}},
  \href{https://doi.org/10.1051/0004-6361/201935641}{\emph{Astron. Astrophys.}
  {\bfseries 629} (2019) A86}
  [\href{https://arxiv.org/abs/1904.03430}{{\ttfamily 1904.03430}}].

\bibitem{Cuceu:2019for}
A.~Cuceu, J.~Farr, P.~Lemos and A.~Font-Ribera, \emph{{Baryon Acoustic
  Oscillations and the Hubble Constant: Past, Present and Future}},
  \href{https://doi.org/10.1088/1475-7516/2019/10/044}{\emph{JCAP} {\bfseries
  10} (2019) 044} [\href{https://arxiv.org/abs/1906.11628}{{\ttfamily
  1906.11628}}].

\bibitem{Schoneberg:2019wmt}
N.~Sch\"oneberg, J.~Lesgourgues and D.C.~Hooper, \emph{{The BAO+BBN take on the
  Hubble tension}},
  \href{https://doi.org/10.1088/1475-7516/2019/10/029}{\emph{JCAP} {\bfseries
  10} (2019) 029} [\href{https://arxiv.org/abs/1907.11594}{{\ttfamily
  1907.11594}}].

\bibitem{Schoneberg:2022ggi}
N.~Sch\"oneberg, L.~Verde, H.~Gil-Mar\'\i{}n and S.~Brieden, \emph{{BAO+BBN
  revisited \textemdash{} growing the Hubble tension with a 0.7 km/s/Mpc
  constraint}},
  \href{https://doi.org/10.1088/1475-7516/2022/11/039}{\emph{JCAP} {\bfseries
  11} (2022) 039} [\href{https://arxiv.org/abs/2209.14330}{{\ttfamily
  2209.14330}}].

\bibitem{Ivanov:2019pdj}
M.M.~Ivanov, M.~Simonovi\'c and M.~Zaldarriaga, \emph{{Cosmological Parameters
  from the BOSS Galaxy Power Spectrum}},
  \href{https://doi.org/10.1088/1475-7516/2020/05/042}{\emph{JCAP} {\bfseries
  05} (2020) 042} [\href{https://arxiv.org/abs/1909.05277}{{\ttfamily
  1909.05277}}].

\bibitem{DAmico:2019fhj}
G.~D'Amico, J.~Gleyzes, N.~Kokron, K.~Markovic, L.~Senatore, P.~Zhang et~al.,
  \emph{{The Cosmological Analysis of the SDSS/BOSS data from the Effective
  Field Theory of Large-Scale Structure}},
  \href{https://doi.org/10.1088/1475-7516/2020/05/005}{\emph{JCAP} {\bfseries
  05} (2020) 005} [\href{https://arxiv.org/abs/1909.05271}{{\ttfamily
  1909.05271}}].

\bibitem{Philcox:2020vvt}
O.H.E.~Philcox, M.M.~Ivanov, M.~Simonovi\'c and M.~Zaldarriaga,
  \emph{{Combining Full-Shape and BAO Analyses of Galaxy Power Spectra: A
  1.6\textbackslash{}\% CMB-independent constraint on H$_0$}},
  \href{https://doi.org/10.1088/1475-7516/2020/05/032}{\emph{JCAP} {\bfseries
  05} (2020) 032} [\href{https://arxiv.org/abs/2002.04035}{{\ttfamily
  2002.04035}}].

\bibitem{Chudaykin:2020aoj}
A.~Chudaykin, M.M.~Ivanov, O.H.E.~Philcox and M.~Simonovi\'c, \emph{{Nonlinear
  perturbation theory extension of the Boltzmann code CLASS}},
  \href{https://doi.org/10.1103/PhysRevD.102.063533}{\emph{Phys. Rev. D}
  {\bfseries 102} (2020) 063533}
  [\href{https://arxiv.org/abs/2004.10607}{{\ttfamily 2004.10607}}].

\bibitem{Wadekar:2020hax}
D.~Wadekar, M.M.~Ivanov and R.~Scoccimarro, \emph{{Cosmological constraints
  from BOSS with analytic covariance matrices}},
  \href{https://doi.org/10.1103/PhysRevD.102.123521}{\emph{Phys. Rev. D}
  {\bfseries 102} (2020) 123521}
  [\href{https://arxiv.org/abs/2009.00622}{{\ttfamily 2009.00622}}].

\bibitem{Kobayashi:2021oud}
Y.~Kobayashi, T.~Nishimichi, M.~Takada and H.~Miyatake, \emph{{Full-shape
  cosmology analysis of the SDSS-III BOSS galaxy power spectrum using an
  emulator-based halo model: A 5\% determination of \ensuremath{\sigma}8}},
  \href{https://doi.org/10.1103/PhysRevD.105.083517}{\emph{Phys. Rev. D}
  {\bfseries 105} (2022) 083517}
  [\href{https://arxiv.org/abs/2110.06969}{{\ttfamily 2110.06969}}].

\bibitem{Chen:2021wdi}
S.-F.~Chen, Z.~Vlah and M.~White, \emph{{A new analysis of galaxy 2-point
  functions in the BOSS survey, including full-shape information and
  post-reconstruction BAO}},
  \href{https://doi.org/10.1088/1475-7516/2022/02/008}{\emph{JCAP} {\bfseries
  02} (2022) 008} [\href{https://arxiv.org/abs/2110.05530}{{\ttfamily
  2110.05530}}].

\bibitem{Smith:2022iax}
T.L.~Smith, V.~Poulin and T.~Simon, \emph{{Assessing the robustness of sound
  horizon-free determinations of the Hubble constant}},
  \href{https://arxiv.org/abs/2208.12992}{{\ttfamily 2208.12992}}.

\bibitem{Simon:2022lde}
T.~Simon, P.~Zhang, V.~Poulin and T.L.~Smith, \emph{{Consistency of effective
  field theory analyses of the BOSS power spectrum}},
  \href{https://doi.org/10.1103/PhysRevD.107.123530}{\emph{Phys. Rev. D}
  {\bfseries 107} (2023) 123530}
  [\href{https://arxiv.org/abs/2208.05929}{{\ttfamily 2208.05929}}].

\bibitem{Holm:2023laa}
E.B.~Holm, L.~Herold, T.~Simon, E.G.M.~Ferreira, S.~Hannestad, V.~Poulin
  et~al., \emph{{Bayesian and frequentist investigation of prior effects in EFT
  of LSS analyses of full-shape BOSS and eBOSS data}},
  \href{https://doi.org/10.1103/PhysRevD.108.123514}{\emph{Phys. Rev. D}
  {\bfseries 108} (2023) 123514}
  [\href{https://arxiv.org/abs/2309.04468}{{\ttfamily 2309.04468}}].

\bibitem{Brieden:2021cfg}
S.~Brieden, H.~Gil-Mar\'\i{}n and L.~Verde, \emph{{Model-independent versus
  model-dependent interpretation of the SDSS-III BOSS power spectrum: Bridging
  the divide}}, \href{https://doi.org/10.1103/PhysRevD.104.L121301}{\emph{Phys.
  Rev. D} {\bfseries 104} (2021) L121301}
  [\href{https://arxiv.org/abs/2106.11931}{{\ttfamily 2106.11931}}].

\bibitem{Brieden:2021edu}
S.~Brieden, H.~Gil-Mar\'\i{}n and L.~Verde, \emph{{ShapeFit: extracting the
  power spectrum shape information in galaxy surveys beyond BAO and RSD}},
  \href{https://doi.org/10.1088/1475-7516/2021/12/054}{\emph{JCAP} {\bfseries
  12} (2021) 054} [\href{https://arxiv.org/abs/2106.07641}{{\ttfamily
  2106.07641}}].

\bibitem{Brieden:2022lsd}
S.~Brieden, H.~Gil-Mar\'\i{}n and L.~Verde, \emph{{Model-agnostic
  interpretation of 10 billion years of cosmic evolution traced by BOSS and
  eBOSS data}},
  \href{https://doi.org/10.1088/1475-7516/2022/08/024}{\emph{JCAP} {\bfseries
  08} (2022) 024} [\href{https://arxiv.org/abs/2204.11868}{{\ttfamily
  2204.11868}}].

\bibitem{Grohs:2023voo}
E.~Grohs and G.M.~Fuller, \emph{{Big Bang Nucleosynthesis}},  in
  \emph{{Handbook of Nuclear Physics}}, I.~Tanihata, H.~Toki and T.~Kajino,
  eds., pp.~1--21 (2023),
  \href{https://doi.org/10.1007/978-981-15-8818-1_127-1}{DOI}
  [\href{https://arxiv.org/abs/2301.12299}{{\ttfamily 2301.12299}}].

\bibitem{Cyburt:2015mya}
R.H.~Cyburt, B.D.~Fields, K.A.~Olive and T.-H.~Yeh, \emph{{Big Bang
  Nucleosynthesis: 2015}},
  \href{https://doi.org/10.1103/RevModPhys.88.015004}{\emph{Rev. Mod. Phys.}
  {\bfseries 88} (2016) 015004}
  [\href{https://arxiv.org/abs/1505.01076}{{\ttfamily 1505.01076}}].

\bibitem{Pitrou:2018cgg}
C.~Pitrou, A.~Coc, J.-P.~Uzan and E.~Vangioni, \emph{{Precision big bang
  nucleosynthesis with improved Helium-4 predictions}},
  \href{https://doi.org/10.1016/j.physrep.2018.04.005}{\emph{Phys. Rept.}
  {\bfseries 754} (2018) 1} [\href{https://arxiv.org/abs/1801.08023}{{\ttfamily
  1801.08023}}].

\bibitem{Consiglio:2017pot}
R.~Consiglio, P.F.~de~Salas, G.~Mangano, G.~Miele, S.~Pastor and O.~Pisanti,
  \emph{{PArthENoPE reloaded}},
  \href{https://doi.org/10.1016/j.cpc.2018.06.022}{\emph{Comput. Phys. Commun.}
  {\bfseries 233} (2018) 237}
  [\href{https://arxiv.org/abs/1712.04378}{{\ttfamily 1712.04378}}].

\bibitem{ParticleDataGroup:2022pth}
{\scshape Particle Data Group} collaboration, \emph{{Review of Particle
  Physics}}, \href{https://doi.org/10.1093/ptep/ptac097}{\emph{PTEP} {\bfseries
  2022} (2022) 083C01}.

\bibitem{Matsumoto:2022tlr}
A.~Matsumoto et~al., \emph{{EMPRESS. VIII. A New Determination of Primordial He
  Abundance with Extremely Metal-poor Galaxies: A Suggestion of the Lepton
  Asymmetry and Implications for the Hubble Tension}},
  \href{https://doi.org/10.3847/1538-4357/ac9ea1}{\emph{Astrophys. J.}
  {\bfseries 941} (2022) 167}
  [\href{https://arxiv.org/abs/2203.09617}{{\ttfamily 2203.09617}}].

\bibitem{Cooke:2017cwo}
R.J.~Cooke, M.~Pettini and C.C.~Steidel, \emph{{One Percent Determination of
  the Primordial Deuterium Abundance}},
  \href{https://doi.org/10.3847/1538-4357/aaab53}{\emph{Astrophys. J.}
  {\bfseries 855} (2018) 102}
  [\href{https://arxiv.org/abs/1710.11129}{{\ttfamily 1710.11129}}].

\bibitem{Aver:2015iza}
E.~Aver, K.A.~Olive and E.D.~Skillman, \emph{{The effects of He I
  \ensuremath{\lambda}10830 on helium abundance determinations}},
  \href{https://doi.org/10.1088/1475-7516/2015/07/011}{\emph{JCAP} {\bfseries
  07} (2015) 011} [\href{https://arxiv.org/abs/1503.08146}{{\ttfamily
  1503.08146}}].

\bibitem{Yeh:2022heq}
T.-H.~Yeh, J.~Shelton, K.A.~Olive and B.D.~Fields, \emph{{Probing physics
  beyond the standard model: limits from BBN and the CMB independently and
  combined}}, \href{https://doi.org/10.1088/1475-7516/2022/10/046}{\emph{JCAP}
  {\bfseries 10} (2022) 046}
  [\href{https://arxiv.org/abs/2207.13133}{{\ttfamily 2207.13133}}].

\bibitem{2022MNRAS.510..373A}
E.~{Aver}, D.A.~{Berg}, A.S.~{Hirschauer}, K.A.~{Olive}, R.W.~{Pogge},
  N.S.J.~{Rogers} et~al., \emph{{A comprehensive chemical abundance analysis of
  the extremely metal poor Leoncino Dwarf galaxy (AGC 198691)}},
  \href{https://doi.org/10.1093/mnras/stab3226}{\emph{\mnras} {\bfseries 510}
  (2022) 373} [\href{https://arxiv.org/abs/2109.00178}{{\ttfamily
  2109.00178}}].

\bibitem{Pisanti:2007hk}
O.~Pisanti, A.~Cirillo, S.~Esposito, F.~Iocco, G.~Mangano, G.~Miele et~al.,
  \emph{{PArthENoPE: Public Algorithm Evaluating the Nucleosynthesis of
  Primordial Elements}},
  \href{https://doi.org/10.1016/j.cpc.2008.02.015}{\emph{Comput. Phys. Commun.}
  {\bfseries 178} (2008) 956}
  [\href{https://arxiv.org/abs/0705.0290}{{\ttfamily 0705.0290}}].

\bibitem{Gariazzo:2021iiu}
S.~{Gariazzo}, P.~{F. de Salas}, O.~{Pisanti} and R.~{Consiglio},
  \emph{{PArthENoPE revolutions}},
  \href{https://doi.org/10.1016/j.cpc.2021.108205}{\emph{Computer Physics
  Communications} {\bfseries 271} (2022) 108205}
  [\href{https://arxiv.org/abs/2103.05027}{{\ttfamily 2103.05027}}].

\bibitem{Olive:1999ij}
K.A.~Olive, G.~Steigman and T.P.~Walker, \emph{{Primordial nucleosynthesis:
  Theory and observations}},
  \href{https://doi.org/10.1016/S0370-1573(00)00031-4}{\emph{Phys. Rept.}
  {\bfseries 333} (2000) 389}
  [\href{https://arxiv.org/abs/astro-ph/9905320}{{\ttfamily
  astro-ph/9905320}}].

\bibitem{1969ApJS...18..247W}
R.V.~{Wagoner}, \emph{{Synthesis of the Elements Within Objects Exploding from
  Very High Temperatures}}, \href{https://doi.org/10.1086/190191}{\emph{\apjs}
  {\bfseries 18} (1969) 247}.

\bibitem{Fields:2019pfx}
B.D.~Fields, K.A.~Olive, T.-H.~Yeh and C.~Young, \emph{{Big-Bang
  Nucleosynthesis after Planck}},
  \href{https://doi.org/10.1088/1475-7516/2020/03/010}{\emph{JCAP} {\bfseries
  03} (2020) 010} [\href{https://arxiv.org/abs/1912.01132}{{\ttfamily
  1912.01132}}].

\bibitem{Arbey:2011nf}
A.~Arbey, \emph{{AlterBBN: A program for calculating the BBN abundances of the
  elements in alternative cosmologies}},
  \href{https://doi.org/10.1016/j.cpc.2012.03.018}{\emph{Comput. Phys. Commun.}
  {\bfseries 183} (2012) 1822}
  [\href{https://arxiv.org/abs/1106.1363}{{\ttfamily 1106.1363}}].

\bibitem{Arbey:2018zfh}
A.~Arbey, J.~Auffinger, K.P.~Hickerson and E.S.~Jenssen, \emph{{AlterBBN v2: A
  public code for calculating Big-Bang nucleosynthesis constraints in
  alternative cosmologies}},
  \href{https://doi.org/10.1016/j.cpc.2019.106982}{\emph{Comput. Phys. Commun.}
  {\bfseries 248} (2020) 106982}
  [\href{https://arxiv.org/abs/1806.11095}{{\ttfamily 1806.11095}}].

\bibitem{Serpico:2004gx}
P.D.~Serpico, S.~Esposito, F.~Iocco, G.~Mangano, G.~Miele and O.~Pisanti,
  \emph{{Nuclear reaction network for primordial nucleosynthesis: A Detailed
  analysis of rates, uncertainties and light nuclei yields}},
  \href{https://doi.org/10.1088/1475-7516/2004/12/010}{\emph{JCAP} {\bfseries
  12} (2004) 010} [\href{https://arxiv.org/abs/astro-ph/0408076}{{\ttfamily
  astro-ph/0408076}}].

\bibitem{Mossa:2020gjc}
V.~Mossa et~al., \emph{{The baryon density of the Universe from an improved
  rate of deuterium burning}},
  \href{https://doi.org/10.1038/s41586-020-2878-4}{\emph{Nature} {\bfseries
  587} (2020) 210}.

\bibitem{Pitrou:2021vqr}
C.~Pitrou, A.~Coc, J.-P.~Uzan and E.~Vangioni, \emph{{Resolving conclusions
  about the early Universe requires accurate nuclear measurements}},
  \href{https://doi.org/10.1038/s42254-021-00294-6}{\emph{Nature Rev. Phys.}
  {\bfseries 3} (2021) 231} [\href{https://arxiv.org/abs/2104.11148}{{\ttfamily
  2104.11148}}].

\bibitem{Burns:2023sgx}
A.-K.~Burns, T.M.P.~Tait and M.~Valli, \emph{{PRyMordial: The First Three
  Minutes, Within and Beyond the Standard Model}},
  \href{https://arxiv.org/abs/2307.07061}{{\ttfamily 2307.07061}}.

\bibitem{2022CoPhC.27108205G}
S.~{Gariazzo}, P.~{F. de Salas}, O.~{Pisanti} and R.~{Consiglio},
  \emph{{PArthENoPE revolutions}},
  \href{https://doi.org/10.1016/j.cpc.2021.108205}{\emph{Computer Physics
  Communications} {\bfseries 271} (2022) 108205}
  [\href{https://arxiv.org/abs/2103.05027}{{\ttfamily 2103.05027}}].

\bibitem{2014ApJ...785...96T}
A.~{Tumino}, R.~{Spart{\`a}}, C.~{Spitaleri}, A.M.~{Mukhamedzhanov},
  S.~{Typel}, R.G.~{Pizzone} et~al., \emph{{New Determination of the
  $^{2}$H(d,p)$^{3}$H and $^{2}$H(d,n)$^{3}$He Reaction Rates at Astrophysical
  Energies}}, \href{https://doi.org/10.1088/0004-637X/785/2/96}{\emph{\apj}
  {\bfseries 785} (2014) 96}.

\bibitem{Xu:2013fha}
Y.~Xu, K.~Takahashi, S.~Goriely, M.~Arnould, M.~Ohta and H.~Utsunomiya,
  \emph{{NACRE II: an update of the NACRE compilation of
  charged-particle-induced thermonuclear reaction rates for nuclei with mass
  number $A < 16$}},
  \href{https://doi.org/10.1016/j.nuclphysa.2013.09.007}{\emph{Nucl. Phys. A}
  {\bfseries 918} (2013) 61} [\href{https://arxiv.org/abs/1310.7099}{{\ttfamily
  1310.7099}}].

\bibitem{Descouvemont:2004cw}
P.~Descouvemont, A.~Adahchour, C.~Angulo, A.~Coc and E.~Vangioni-Flam,
  \emph{{Compilation and R-matrix analysis of Big Bang nuclear reaction
  rates}}, \href{https://doi.org/10.1016/j.adt.2004.08.001}{\emph{Atom. Data
  Nucl. Data Tabl.} {\bfseries 88} (2004) 203}
  [\href{https://arxiv.org/abs/astro-ph/0407101}{{\ttfamily
  astro-ph/0407101}}].

\bibitem{Iliadis:2016vkw}
C.~Iliadis, K.~Anderson, A.~Coc, F.~Timmes and S.~Starrfield, \emph{{Bayesian
  Estimation of Thermonuclear Reaction Rates}},
  \href{https://doi.org/10.3847/0004-637X/831/1/107}{\emph{Astrophys. J.}
  {\bfseries 831} (2016) 107}
  [\href{https://arxiv.org/abs/1608.05853}{{\ttfamily 1608.05853}}].

\bibitem{InestaGomez:2017eya}
A.~I\~nesta G\'omez, C.~Iliadis and A.~Coc, \emph{{Bayesian estimation of
  thermonuclear reaction rates for deuterium+deuterium reactions}},
  \href{https://doi.org/10.3847/1538-4357/aa9025}{\emph{Astrophys. J.}
  {\bfseries 849} (2017) 134}
  [\href{https://arxiv.org/abs/1710.01647}{{\ttfamily 1710.01647}}].

\bibitem{2019ApJ...872...75D}
R.S.~{de Souza}, C.~{Iliadis} and A.~{Coc}, \emph{{Astrophysical S-factors,
  Thermonuclear Rates, and Electron Screening Potential for the
  $^{3}$He(d,p)$^{4}$He Big Bang Reaction via a Hierarchical Bayesian Model}},
  \href{https://doi.org/10.3847/1538-4357/aafda9}{\emph{\apj} {\bfseries 872}
  (2019) 75} [\href{https://arxiv.org/abs/1809.06966}{{\ttfamily 1809.06966}}].

\bibitem{deSouza:2019pmr}
R.S.~de~Souza, S.R.~Boston, A.~Coc and C.~Iliadis, \emph{{Thermonuclear fusion
  rates for tritium + deuterium using Bayesian methods}},
  \href{https://doi.org/10.1103/PhysRevC.99.014619}{\emph{Phys. Rev. C}
  {\bfseries 99} (2019) 014619}
  [\href{https://arxiv.org/abs/1901.04857}{{\ttfamily 1901.04857}}].

\bibitem{Moscoso:2021xog}
J.~Moscoso, R.S.~de~Souza, A.~Coc and C.~Iliadis, \emph{{Bayesian Estimation of
  the D(p,\ensuremath{\gamma})$^{3}$He Thermonuclear Reaction Rate}},
  \href{https://doi.org/10.3847/1538-4357/ac1db0}{\emph{Astrophys. J.}
  {\bfseries 923} (2021) 49}
  [\href{https://arxiv.org/abs/2109.00049}{{\ttfamily 2109.00049}}].

\bibitem{UCNt:2021pcg}
{\scshape UCN\ensuremath{\tau}} collaboration, \emph{{Improved neutron lifetime
  measurement with UCN$\tau$}},
  \href{https://doi.org/10.1103/PhysRevLett.127.162501}{\emph{Phys. Rev. Lett.}
  {\bfseries 127} (2021) 162501}
  [\href{https://arxiv.org/abs/2106.10375}{{\ttfamily 2106.10375}}].

\bibitem{PhysRevD.98.030001}
{\scshape Particle Data Group} collaboration, \emph{Review of particle
  physics}, \href{https://doi.org/10.1103/PhysRevD.98.030001}{\emph{Phys. Rev.
  D} {\bfseries 98} (2018) 030001}.

\bibitem{Chowdhury:2022ahn}
T.~Chowdhury and S.~Ipek, \emph{{Neutron Lifetime Anomaly and Big Bang
  Nucleosynthesis}},  \href{https://arxiv.org/abs/2210.12031}{{\ttfamily
  2210.12031}}.

\bibitem{Brinckmann:2018cvx}
T.~Brinckmann and J.~Lesgourgues, \emph{{MontePython 3: boosted MCMC sampler
  and other features}},
  \href{https://doi.org/10.1016/j.dark.2018.100260}{\emph{Phys. Dark Univ.}
  {\bfseries 24} (2019) 100260}
  [\href{https://arxiv.org/abs/1804.07261}{{\ttfamily 1804.07261}}].

\bibitem{Burns:2022hkq}
A.-K.~Burns, T.M.P.~Tait and M.~Valli, \emph{{Indications for a Nonzero Lepton
  Asymmetry from Extremely Metal-Poor Galaxies}},
  \href{https://doi.org/10.1103/PhysRevLett.130.131001}{\emph{Phys. Rev. Lett.}
  {\bfseries 130} (2023) 131001}
  [\href{https://arxiv.org/abs/2206.00693}{{\ttfamily 2206.00693}}].

\bibitem{Takahashi:2022cpn}
T.~Takahashi and S.~Yamashita, \emph{{Big bang nucleosynthesis and early dark
  energy in light of the EMPRESS Yp results and the H0 tension}},
  \href{https://doi.org/10.1103/PhysRevD.107.103520}{\emph{Phys. Rev. D}
  {\bfseries 107} (2023) 103520}
  [\href{https://arxiv.org/abs/2211.04087}{{\ttfamily 2211.04087}}].

\bibitem{Aver:2020fon}
E.~Aver, D.A.~Berg, K.A.~Olive, R.W.~Pogge, J.J.~Salzer and E.D.~Skillman,
  \emph{{Improving helium abundance determinations with Leo P as a case
  study}}, \href{https://doi.org/10.1088/1475-7516/2021/03/027}{\emph{JCAP}
  {\bfseries 03} (2021) 027}
  [\href{https://arxiv.org/abs/2010.04180}{{\ttfamily 2010.04180}}].

\bibitem{Valerdi:2019beb}
M.~Valerdi, A.~Peimbert, M.~Peimbert and A.~Sixtos, \emph{{Determination of the
  Primordial Helium Abundance Based on NGC 346, an H ii Region of the Small
  Magellanic Cloud}},
  \href{https://doi.org/10.3847/1538-4357/ab14e4}{\emph{Astrophys. J.}
  {\bfseries 876} (2019) 98}
  [\href{https://arxiv.org/abs/1904.01594}{{\ttfamily 1904.01594}}].

\bibitem{Fernandez:2019hds}
V.~Fern\'andez, E.~Terlevich, A.I.~D\'\i{}az and R.~Terlevich, \emph{{A
  Bayesian direct method implementation to fit emission line spectra:
  Application to the primordial He abundance determination}},
  \href{https://doi.org/10.1093/mnras/stz1433}{\emph{Mon. Not. Roy. Astron.
  Soc.} {\bfseries 487} (2019) 3221}
  [\href{https://arxiv.org/abs/1905.09215}{{\ttfamily 1905.09215}}].

\bibitem{2020ApJ...896...77H}
T.~{Hsyu}, R.J.~{Cooke}, J.X.~{Prochaska} and M.~{Bolte}, \emph{{The PHLEK
  Survey: A New Determination of the Primordial Helium Abundance}},
  \href{https://doi.org/10.3847/1538-4357/ab91af}{\emph{\apj} {\bfseries 896}
  (2020) 77} [\href{https://arxiv.org/abs/2005.12290}{{\ttfamily 2005.12290}}].

\bibitem{Kurichin:2021ppm}
O.A.~Kurichin, P.A.~Kislitsyn, V.V.~Klimenko, S.A.~Balashev and A.V.~Ivanchik,
  \emph{{A new determination of the primordial helium abundance using the
  analyses of H II region spectra from SDSS}},
  \href{https://doi.org/10.1093/mnras/stab215}{\emph{Mon. Not. Roy. Astron.
  Soc.} {\bfseries 502} (2021) 3045}
  [\href{https://arxiv.org/abs/2101.09127}{{\ttfamily 2101.09127}}].

\bibitem{2021MNRAS.505.3624V}
M.~{Valerdi}, A.~{Peimbert} and M.~{Peimbert}, \emph{{Chemical abundances in
  seven metal-poor H II regions and a determination of the primordial helium
  abundance}}, \href{https://doi.org/10.1093/mnras/stab1543}{\emph{\mnras}
  {\bfseries 505} (2021) 3624}
  [\href{https://arxiv.org/abs/2105.12260}{{\ttfamily 2105.12260}}].

\bibitem{Pettini:2001yu}
M.~Pettini and D.V.~Bowen, \emph{{A new measurement of the primordial abundance
  of deuterium: toward convergence with the baryon density from the cmb?}},
  \href{https://doi.org/10.1086/322510}{\emph{Astrophys. J.} {\bfseries 560}
  (2001) 41} [\href{https://arxiv.org/abs/astro-ph/0104474}{{\ttfamily
  astro-ph/0104474}}].

\bibitem{DOdorico:2001vgo}
S.~D'Odorico, M.~Dessauges-Zavadsky and P.~Molaro, \emph{{A new deuterium
  abundance measurement from a damped ly-alpha system at z\_abs = 3.025}},
  \href{https://doi.org/10.1051/0004-6361:20010183}{\emph{Astron. Astrophys.}
  {\bfseries 368} (2001) L21}
  [\href{https://arxiv.org/abs/astro-ph/0102162}{{\ttfamily
  astro-ph/0102162}}].

\bibitem{Srianand:2010un}
R.~Srianand, N.~Gupta, P.~Petitjean, P.~Noterdaeme and C.~Ledoux,
  \emph{{Detection of 21-cm, H2 and Deuterium absorption at z\ensuremath{>}3
  along the line-of-sight to J1337+3152}},
  \href{https://doi.org/10.1111/j.1365-2966.2010.16574.x}{\emph{Mon. Not. Roy.
  Astron. Soc.} {\bfseries 405} (2010) 1888}
  [\href{https://arxiv.org/abs/1002.4620}{{\ttfamily 1002.4620}}].

\bibitem{Fumagalli:2011iw}
M.~Fumagalli, J.M.~O'Meara and J.X.~Prochaska, \emph{{Detection of Pristine Gas
  Two Billion Years after the Big Bang}},
  \href{https://doi.org/10.1126/science.1213581}{\emph{Science} {\bfseries 334}
  (2011) 1245} [\href{https://arxiv.org/abs/1111.2334}{{\ttfamily 1111.2334}}].

\bibitem{Noterdaeme:2012pa}
P.~Noterdaeme, S.~Lopez, V.~Dumont, C.~Ledoux, P.~Molaro and P.~Petitjean,
  \emph{{Deuterium at high-redshift: Primordial abundance in the zabs = 2.621
  damped Ly-alpha system towards CTQ247}},
  \href{https://doi.org/10.1051/0004-6361/201219453}{\emph{Astron. Astrophys.}
  {\bfseries 542} (2012) L33}
  [\href{https://arxiv.org/abs/1205.3777}{{\ttfamily 1205.3777}}].

\bibitem{Cooke:2013cba}
R.~Cooke, M.~Pettini, R.A.~Jorgenson, M.T.~Murphy and C.C.~Steidel,
  \emph{{Precision measures of the primordial abundance of deuterium}},
  \href{https://doi.org/10.1088/0004-637X/781/1/31}{\emph{Astrophys. J.}
  {\bfseries 781} (2014) 31} [\href{https://arxiv.org/abs/1308.3240}{{\ttfamily
  1308.3240}}].

\bibitem{Riemer-Sorensen:2014aoa}
S.~Riemer-S\o{}rensen, J.K.~Webb, N.~Crighton, V.~Dumont, K.~Ali, S.~Kotu\v{s}
  et~al., \emph{{A robust deuterium abundance; Re-measurement of the z=3.256
  absorption system towards the quasar PKS1937-1009}},
  \href{https://doi.org/10.1093/mnras/stu2599}{\emph{Mon. Not. Roy. Astron.
  Soc.} {\bfseries 447} (2015) 2925}
  [\href{https://arxiv.org/abs/1412.4043}{{\ttfamily 1412.4043}}].

\bibitem{Balashev:2015hoe}
S.A.~Balashev, E.O.~Zavarygin, A.V.~Ivanchik, K.N.~Telikova and
  D.A.~Varshalovich, \emph{{The primordial deuterium abundance: subDLA system
  at $z_{\rm abs}=2.437$ towards the QSO J 1444+2919}},
  \href{https://doi.org/10.1093/mnras/stw356}{\emph{Mon. Not. Roy. Astron.
  Soc.} {\bfseries 458} (2016) 2188}
  [\href{https://arxiv.org/abs/1511.01797}{{\ttfamily 1511.01797}}].

\bibitem{Riemer-Sorensen:2017pey}
S.~Riemer-S\o{}rensen, S.~Kotu\v{s}, J.K.~Webb, K.~Ali, V.~Dumont, M.T.~Murphy
  et~al., \emph{{A precise deuterium abundance: remeasurement of the z = 3.572
  absorption system towards the quasar PKS1937\textendash{}101}},
  \href{https://doi.org/10.1093/mnras/stx681}{\emph{Mon. Not. Roy. Astron.
  Soc.} {\bfseries 468} (2017) 3239}
  [\href{https://arxiv.org/abs/1703.06656}{{\ttfamily 1703.06656}}].

\bibitem{Zavarygin:2017cov}
E.O.~Zavarygin, J.K.~Webb, V.~Dumont and S.~Riemer-S\o{}rensen, \emph{{The
  primordial deuterium abundance at zabs~=~2.504 from a high signal-to-noise
  spectrum of Q1009+2956}},
  \href{https://doi.org/10.1093/mnras/sty1003}{\emph{Mon. Not. Roy. Astron.
  Soc.} {\bfseries 477} (2018) 5536}
  [\href{https://arxiv.org/abs/1706.09512}{{\ttfamily 1706.09512}}].

\bibitem{2012MNRAS.425.2477P}
M.~{Pettini} and R.~{Cooke}, \emph{{A new, precise measurement of the
  primordial abundance of deuterium}},
  \href{https://doi.org/10.1111/j.1365-2966.2012.21665.x}{\emph{\mnras}
  {\bfseries 425} (2012) 2477}
  [\href{https://arxiv.org/abs/1205.3785}{{\ttfamily 1205.3785}}].

\bibitem{2018MNRAS.477.5536Z}
E.O.~{Zavarygin}, J.K.~{Webb}, V.~{Dumont} and S.~{Riemer-S{\o}rensen},
  \emph{{The primordial deuterium abundance at z$_{abs}$ = 2.504 from a high
  signal-to-noise spectrum of Q1009+2956}},
  \href{https://doi.org/10.1093/mnras/sty1003}{\emph{\mnras} {\bfseries 477}
  (2018) 5536} [\href{https://arxiv.org/abs/1706.09512}{{\ttfamily
  1706.09512}}].

\bibitem{Cooke:2016rky}
R.J.~Cooke, M.~Pettini, K.M.~Nollett and R.~Jorgenson, \emph{{The primordial
  deuterium abundance of the most metal-poor damped Ly$\alpha$ system}},
  \href{https://doi.org/10.3847/0004-637X/830/2/148}{\emph{Astrophys. J.}
  {\bfseries 830} (2016) 148}
  [\href{https://arxiv.org/abs/1607.03900}{{\ttfamily 1607.03900}}].

\end{thebibliography}\endgroup
\bibliographystyle{JHEP}

\end{document}